\documentclass[a4paper,11pt]{article}
\pdfoutput=1 

\usepackage{jcappub} 

\usepackage[T1]{fontenc} 
\usepackage{orcidlink}
\usepackage{graphicx}
\usepackage{amsfonts}
\usepackage{amssymb}
\usepackage{amsbsy}
\usepackage{amsmath}
\usepackage{mathtools}
\usepackage[normalem]{ulem}
\usepackage{latexsym}
\usepackage{bm}
\usepackage{color}
\usepackage{hyperref}
\usepackage[top=2cm,bottom=2.5cm,left=2cm,right=2cm]{geometry}
\usepackage{comment}
\usepackage[makeroom]{cancel}
\usepackage{subcaption}
\usepackage{placeins}

\usepackage{stackengine}
\stackMath
\newcommand\tenq[2][1]{%
 \def\useanchorwidth{T}%
  \ifnum#1>1%
    \stackon[0pt]{\tenq[\Lambdamexpr#1-1\relax]{#2}}{\scriptscriptstyle\sim}%
  \else%
    \stackon[1pt]{#2}{\scriptscriptstyle\sim}%
  \fi%
}

\newcommand{\be}{\begin{equation}}
\newcommand{\ee}{\end{equation}}
\newcommand{\de}{\mbox{d}}

\newcommand{\lf}{\left}
\newcommand{\rg}{\right}

\numberwithin{equation}{section}
\renewcommand{\theequation}{\arabic{section}.\arabic{equation}}

\makeatletter 
\gdef\@fpheader{}
\makeatother

\begin{document}


\title{Multiple-scale analysis of modified gravitational-wave propagation}
\footnotesize\texttt{KCL-PH-TH-2025-25}

\author[a,b]{Marco de Cesare}
\author[c]{Mairi Sakellariadou}
\author[c]{and Benjamin Sutton}

\affiliation[a]{Scuola Superiore Meridionale, \\
Largo San Marcellino 10, 80138 Napoli, Italy}
\affiliation[b]{INFN, Sezione di Napoli,\\ Monte S. Angelo, Via Cintia, 80126 Napoli, Italy}
\affiliation[c]{Theoretical Particle Physics and Cosmology Group, Physics Department,
King’s College London, University of London, \\ Strand, London WC2R 2LS, United Kingdom}

\emailAdd{marco.decesare@na.infn.it}
\emailAdd{mairi.sakellariadou@kcl.ac.uk}
\emailAdd{benjamin.j.sutton@kcl.ac.uk}

\abstract{
We employ multiple-scale analysis to systematically derive analytical approximations describing the cosmological propagation of gravitational waves beyond general relativity, in a framework with two interacting spin-2 fields with time-dependent couplings. Such techniques allow us to accurately track the evolution of a system with slowly evolving time-dependent couplings over a large number of oscillation periods.
We focus on tensor modes propagating on sub-horizon scales in a universe dominated by dark energy and explicitly derive solutions for a general class of models.
To illustrate the possible applications of our general scheme and further corroborate our analytical results, we calculate the evolution of tensor perturbations in some phenomenological toy models and compare them with numerical simulations.
We show that, generically, the interactions of independent spin-2 fields lead to non-trivial modifications to the amplitude and phase of the detected waveform, which are different from those obtained in other modified gravity theories with a single graviton. This provides an avenue to test and constrain gravitational models with new fundamental physical fields.
}

\maketitle


\section{Introduction}


Since the advent of gravitational wave (GW) astronomy, beginning with the first detection of a binary black hole (BBH) merger \cite{LIGOScientific:2016aoc}, there have been $\mathcal{O}(100)$ GW detections to date. By the completion of the fifth observing run (O5) of LIGO this is expected to reach thousands of detections \cite{KAGRA:2013rdx}. With this wealth of data at our disposal, it has become increasingly prudent to test the fundamental physics of gravity, and constrain deviations from general relativity (GR) with observations. In perturbative GR to linear order, there is one dynamical spin-2 field propagating which carries two polarizations. However, in general, in metric theories of gravity, there are up to six possible polarizations of GWs. In a homogeneous background, GWs do not couple to scalar or vector modes to linear order in perturbation theory (this is not necessarily true for inhomogeneous \cite{Garoffolo:2019mna} or anisotropic \cite{Pereira:2007yy} backgrounds). Therefore, it is prudent to assume the exotic sector is described by another spin-2 field coupled to the usual tensorial perturbation of GR, in models where gravity is coupled to another cosmological field. There is a landscape of models where an additional propagating helicity-2 partner arises, such as Hassan-Rosen bimetric gravity \cite{Hassan:2011hr,Hassan:2011zd}, Yang-Mills theories \cite{Cervero:1978db, Galtsov:1991un,Darian:1996mb}, Abelian multi-gauge fields in a gaugid configuration \cite{Piazza:2017bsd} and multi-Proca theories \cite{Armendariz-Picon:2004say, Hull:2014bga, Allys:2016kbq, BeltranJimenez:2016afo}. Another class of models which give rise to a second tensor mode occurs when a cosmological model contains fields whose vacuum expectation values violate homogeneity and/or isotropy \cite{Maleknejad:2011jw}. Although these fields are a priori in disagreement with the cosmological principle, a homogeneous and isotropic universe may still be recovered if the fields have internal symmetries which hide the apparent violation of the cosmological principle. Perturbations on this non-trivial background configuration may then arrange themselves into a helicity-2 partner without the theory needing an explicit spin-2 field \cite{BeltranJimenez:2019xxx}. This scenario has been explored in models with vector fields in Refs.~\cite{Maleknejad:2011jw, Adshead:2012kp, Maleknejad:2012fw, BeltranJimenez:2018ymu, Piazza:2017bsd}.

A model-agnostic, purely phenomenological approach to analyze the dynamics of GWs coupled to an additional spin-2 field was initially outlined in Ref.~\cite{BeltranJimenez:2019xxx}. Without specifying a 
modified gravity theory, we can describe the propagation of two coupled tensorial perturbations, $h$ and $s$, by the following equation (omitting polarization labels)
\begin{equation}\label{Eq:SecondOrderEoM}
    \bigg[\hat{I}\frac{\de^2}{\de\eta^2}+\hat{\Lambda}(\eta)\frac{\de}{\de\eta}+\hat{C}(\eta)k^2+\hat{\Pi}(\eta)k+\hat{M}(\eta)\bigg]\begin{pmatrix} h(\eta) \\ s(\eta) \end{pmatrix}=0~,
\end{equation} 
where $\hat{\Lambda}$ is the friction matrix, $\hat{C}$ is the velocity matrix, $\hat{\Pi}$ is the chiral matrix, and $\hat{M}$ is the mass matrix. 
For a given model descended from a variational principle, the entries of these matrices may be determined by direct comparison with the perturbative field equations around a cosmological background.
However, in this paper we do not restrict our analysis to a particular theory, but rather work only at the level of the dynamical equation \eqref{Eq:SecondOrderEoM}. 
In this framework, it is also assumed that only the perturbation $h$ couples to matter.
As such, observationally we would not directly detect the exotic mode $s$ with GW detectors; rather, its existence may only be inferred indirectly, as interactions between the two tensor modes lead to deviations in the waveform $h$ predicted by GR. The deviation of this $h$ signal from GR has a rich phenomenology based on the model considered. 
The phenomenology of this class of models featuring two interacting tensor fields was further explored in Refs.~\cite{Ezquiaga:2021ler,Ezquiaga:2022nak}, where, likewise with the original work \cite{BeltranJimenez:2019xxx}, all approximate solutions were obtained using a modified WKB approach. 
However, it has been shown that such a generalization of the WKB method to coupled oscillator systems is mathematically ill-posed and may be limited in its application, and over sufficiently long timescales the approximations may no longer be valid, as discussed in Ref.~\cite{Brizuela:2023uwt}.

Here, under the assumption that the matrices in Eq.~\eqref{Eq:SecondOrderEoM} are slowly varying functions of time, we derive a set of general approximate solutions to this system using multiple-scale expansion techniques, following the general scheme presented in Ref.~\cite{kevorkian_multiple_1996}. 
This method stresses the evolution of systems involving different timescales (organised hierarchically), and ensures uniformly valid approximations by eliminating secular terms at each order in the perturbative expansion. 
The same techniques have been applied previously in the context of bimetric gravity in Ref.~\cite{Brizuela:2023uwt}, while in this work we consider a broader class of models.\footnote{In a different context, multiple-scale analysis has found application in gravitational physics also in the study of extreme mass-ratio inspirals (EMRIs) in the so-called post-adiabatic approximation \cite{Barack:2018yvs}.}
The purpose of this work is not to explore the rich phenomenology of these systems, but to provide the methodology to obtain analytic approximations describing the evolution of tensorial modes.

The paper is organized as follows. In Section~\ref{Sec:interacting_tensors}, we recast the dynamical equations for tensor modes $h$ and $s$ in a form that is better suited for the analysis of sub-horizon scales, introducing an analogue of the standard Mukhanov-Sasaki variables. In Section~\ref{Methodology}, after recasting the dynamics in first-order form in terms of action-angle variables and identifying a characteristic small parameter $\epsilon$, we proceed systematically to a multiple-scale expansion of the system and derive approximate analytical solutions to first order in $\epsilon$. In Section \ref{Sec:pheno_model}, as an illustrative example of the potential applications of the method, we then apply our general results to some particular toy models, including a model with a non-trivial time dependence and mixing. In Section~\ref{Conclusions}, we discuss key results and future outlook of this work. Three technical appendices are also included, which contain more details on the solution in the general case and for the toy models considered.

\section{Interacting tensor modes in de Sitter}\label{Sec:interacting_tensors}

We wish to analyze the dynamics of
coupled helicity-2 modes described by Eq.~\eqref{Eq:SecondOrderEoM} on sub-horizon scales, during the late-time epoch of dark energy (DE) domination. For simplicity, we model dark energy as a cosmological constant. Thus, the cosmological background we are considering is de Sitter spacetime, and the scale factor evolves inversely with conformal time, $a=-(H\eta)^{-1}$, where $H$ is the Hubble constant. The conformal Hubble rate evolves as $\mathcal{H}=-\eta^{-1}$, and $\eta < 0$ where $\eta \rightarrow 0$ corresponds to the asymptotic future. 
We analyze this class of models in the limit where interactions between the two tensor fields are small, as we expect deviations from GR to be small.
Despite remaining close to the limit of GR, the phenomenology of the modified waveform to the perturbation $h$
due to interactions
with $s$ remains non-trivial and rich. We then proceed to manipulate our equations of motion Eq.~\eqref{Eq:SecondOrderEoM} under these conditions. We parameterize the friction matrix as
\begin{equation}
    \hat{\Lambda}(\eta)= \begin{pmatrix}
        \Lambda_{11} (\eta) & \Lambda_{12} (\eta) \\ \Lambda_{21}
        (\eta) & \Lambda_{22} (\eta)
    \end{pmatrix}~.
\end{equation}
Similar notational conventions also apply for the matrix elements of $\hat{C}$, $\hat{\Pi}$, $\hat{M}$.
We also assume that the entries of the friction matrix are described by a Laurent series starting with a term $\mathcal{O}(\eta^{-1})$. A similar parameterization of the friction matrix has also been considered in the literature  \cite{LISACosmologyWorkingGroup:2019mwx,Saltas:2014dha, Nishizawa:2017nef, Belgacem:2017ihm,Lombriser:2015sxa}, and the standard friction term must be recovered in the GR limit.  


It is convenient to rescale the fields in analogy with the definition of the standard Mukhanov-Sasaki variables. This can be achieved by identifying suitable integrating factors, so as to remove the friction terms and work with a system of undamped oscillators with time-dependent frequencies and forcing terms.
Specifically, we introduce the following rescaled variables for the tensor modes
\begin{subequations}\label{Eq:changevariables_munu}
\begin{align}
    &\mu\equiv \exp\left(\int\de\eta\; \frac{\Lambda_{11}}{2} \right)h ~, \\[2mm]
%
    &\nu\equiv\exp\left(\int\de\eta\;\frac{\Lambda_{22}}{2}\right)s~. 
\end{align}
\end{subequations}
For convenience, we define
\be
P_{ij} \equiv C_{ij}+\frac{1}{k}\Pi_{ij}+\frac{1}{k^2}M_{ij}
\ee
to make the equations more compact. Henceforth, we also omit the explicit time dependence to make the notation lighter. With the above, Eq.~\eqref{Eq:SecondOrderEoM} is recast in the following form

\begin{subequations}\label{Eq:Eq:SecondOrderEoM_MSform0}
\begin{align}
\begin{split}
        &\frac{\de^2\mu}{\de\eta^2}+\left(k^2 P_{11}-\frac{1}{2}\Lambda'_{11}-\frac{1}{4}\Lambda^2_{11} \right)\mu \\ &= -\exp\left(\frac{1}{2}\int\de\eta\, \left(\Lambda_{11}-\Lambda_{22} \right)\right)\left[ \Lambda_{12
        }\frac{\de\nu}{\de\eta} +\left(k^2 P_{12}-\frac{1}{2}\Lambda_{12}\Lambda_{22}\right)\nu\right] ~,\\
\end{split}\\[5mm]
\begin{split}
        &\frac{\de^2\nu}{\de\eta^2}+\left(k^2 P_{22}-\frac{1}{2}\Lambda'_{22}-\frac{1}{4}\Lambda^2_{22} \right)\nu \\ &= -\exp\left(\frac{1}{2}\int\de\eta\,\left(\Lambda_{22}-\Lambda_{11}\right) \right)\left[ \Lambda_{21}\frac{\de\mu}{\de\eta} +\left(k^2 P_{21}-\frac{1}{2}\Lambda_{21}\Lambda_{11}\right)\mu\right]~.
\end{split}
\end{align}
\end{subequations}
Next, we expand the entries of the friction matrix as Laurent series with a simple pole at $\eta=0$~, $\Lambda_{ij} = \sum\limits^{+\infty}_{k=-1} \Lambda^{(k)}_{ij}\eta^{k}$~, with constant coefficients $\Lambda^{(k)}_{ij}$.
We treat terms in the expansion of higher order than $\mathcal{O}(\eta^{-1})$ as subleading, which is justified in the late-time ($\eta\to0$) limit.
With this approximation, the system~\eqref{Eq:Eq:SecondOrderEoM_MSform0} simplifies to







%
%
%
\begin{subequations}
    \begin{align}
\begin{split}
        &\frac{\de^2\mu}{\de\eta^2}+\left(k^2 P_{11}+\frac{\Lambda^{(-1)}_{11}(2-\Lambda^{(-1)}_{11})}{4\eta^2}
        \right)\mu \\ 
        &= -\left(\frac{\eta}{\eta_0}\right)^{\frac{1}{2}\left(\Lambda^{(-1)}_{11}-\Lambda^{(-1)}_{22}\right)} \left[ \left(k^2 P_{12}-\frac{\Lambda^{(-1)}_{22}\Lambda^{(-1)}_{12}}{2\eta^2}\right)\nu+\frac{\Lambda^{(-1)}_{12}}{\eta}\frac{\de\nu}{\de\eta} \right]~,\\
\end{split}\\[5mm]
\begin{split}
        &\frac{\de^2\nu}{\de\eta^2}+\left(k^2 P_{22}+\frac{\Lambda^{(-1)}_{22}(2-\Lambda^{(-1)}_{22})}{4\eta^2}
        \right)\nu \\ 
        &= -\left(\frac{\eta}{\eta_0}\right)^{\frac{1}{2}\left(\Lambda^{(-1)}_{22}-\Lambda^{(-1)}_{11}\right)}\left[ \left(k^2 P_{21}-\frac{\Lambda^{(-1)}_{21}\Lambda^{\small{(-1)}}_{11}}{2\eta^2}\right)\mu+\frac{\Lambda^{(-1)}_{21}}{\eta}\frac{\de\mu}{\de\eta} \right]~,
\end{split}
    \end{align}
\end{subequations}
where $\eta_0$ arises as a constant of integration, and is fixed to the value of conformal time at emission.

Since our goal is to study the propagation of sub-horizon modes, the interval of conformal time that we should consider is
\begin{equation}\label{Eq:validity_domain}
    \eta_{\rm\scriptscriptstyle DE} \lesssim \eta_{0} \lesssim \eta \lesssim \eta_{\rm exit} \lesssim 0 ~,
\end{equation}
where $\eta_{\rm exit}$ is the value of conformal time when the GW exits the horizon. (Recall that, in de Sitter space, $\eta < 0$ and the $\eta \rightarrow 0$ limit corresponds to the asymptotic future.)
In the following sections, we will develop approximate analytic solutions in the domain of conformal time specified by \eqref{Eq:validity_domain}. 
Specifically, our approximations will be valid in the regime where $  k|\eta_{0}| \gtrsim k|\eta|\gg1 $~ (that is, till horizon exit, where~$k|\eta_{\rm exit}|\simeq 1$).

In the following, we work under the assumption that the functions $P_{ij}$ are slowly evolving on cosmological timescales of the order $\sim H^{-1}$, which is much larger than the period of a single oscillation $\sim k^{-1}$. Such a hierarchy of timescales enables us to apply multiple-scale analysis to study the dynamics of the system analytically.

\section{Multiple-scale expansion in the sub-horizon regime}\label{Methodology}
To proceed with the application of multiple-scale analysis to the system at hand, 
we start by recasting the system of coupled ODEs \eqref{Eq:SecondOrderEoM} in a dimensionless form. 
For this purpose, we define a dimensionless time coordinate $t \equiv k \eta$~. In this way, we obtain
\begin{subequations}\label{Eq:dimensionless_time_system_pre-epsilon}
\begin{align}
    &\ddot{\mu}+\left(P_{11}+\frac{\Lambda^{(-1)}_{11}(2-\Lambda^{(-1)}_{11})}{4t^2} \right)\mu 
    =-\left(\frac{t}{k\eta_0} \right)^{\frac{1}{2}\left(\Lambda^{(-1)}_{11}-\Lambda^{(-1)}_{22}\right)}\left[ \frac{\Lambda^{(-1)}_{12}}{t}\dot{\nu}+\left(P_{12}-\frac{\Lambda^{(-1)}_{22}\Lambda^{(-1)}_{12}}{2t^2} \right)\nu \right]~,\\[5mm]
    &\ddot{\nu}+\left(P_{22}+\frac{\Lambda^{(-1)}_{22}(2-\Lambda^{(-1)}_{22})}{4t^2} \right)\nu 
    =-\left(\frac{t}{k\eta_0} \right)^{\frac{1}{2}\left(\Lambda^{(-1)}_{22}-\Lambda^{(-1)}_{11}\right)}\left[ \frac{\Lambda^{(-1)}_{21}}{t}\dot{\mu}+\left(P_{21}-\frac{\Lambda^{(-1)}_{11}\Lambda^{(-1)}_{21}}{2t^2} \right)\mu \right]~,
    \end{align}
\end{subequations}
where overdots are used as a shorthand notation for derivatives with respect to $t$~, that is $\dot{\phantom{\mu}}\equiv \frac{\de}{\de t}$~.
Note that, since the $P_{ij}$ are assumed to be slowly evolving on cosmological timescales, they only depend on conformal time through the combination $(k/H)^{-1}\eta$~. Moreover, we have assumed the propagation to be on sub-horizon scales from the time of emission till detection, and therefore the inequality $k|\eta|\ll1$ holds throughout the evolution in the regime of interest.
Then, it is clear that the system \eqref{Eq:dimensionless_time_system_pre-epsilon} naturally involves a small parameter that we define as $\epsilon \equiv 1/k|\eta_0| \ll 1$~.\\
The role played by the small parameter $\epsilon$ is twofold: on the one hand, it governs the magnitude of correction terms to the standard propagation equations, and thus enables us to perturbatively expand the equations of motion; on the other hand, it defines a slow time variable of the system $T \equiv \epsilon\, t$~, which the oscillator couplings depend upon.
In this context, `slow' can be interpreted as scales that are longer than the time of horizon crossing for the modes.
Henceforth, in this notation, our assumptions on $P_{ij}$ imply that these only depend on time through $T$~, that is $P_{ij}=P_{ij}(k,T;\epsilon)$ (where the last entry represents additional parametric dependence on $\epsilon$ besides the combination $T=\epsilon\, t$~).

The purpose of introducing two separate time variables---which are treated as independent in multiple-scale analysis---is so that we may construct a new set of dynamical variables that
behaves approximately as adiabatic invariants, as shown in further detail later in this section.
Further, treating these two time variables as independent in the perturbative expansion provides an additional
degree of freedom which may be exploited to eliminate secular terms at each order in $\epsilon$, and thus return uniform approximations on large timescales of order $t \sim \epsilon^{-1}$~.
Specifically, the scale of the perturbative expansion parameter permits us to robustly define the domain of validity for the approximations in terms of the dimensionless time variable $t$ as
\begin{equation}
    |t_{\text{exit}}|\ll |t| \leq |t_0|  ~,
\end{equation}
where $t_{\text{exit}}=-1$ and initial conditions are set at $t_0=(k\eta_0)$~. 

%

\subsection{Recasting the dynamics in action-angle variables} Equations~\eqref{Eq:dimensionless_time_system_pre-epsilon} are in the form of coupled non-autonomous linear oscillators, generically described by


%
\begin{equation}
    \ddot{x}_N+\gamma_N^2 x_N=\epsilon\,\xi_N~,
\end{equation}
where the label $N$ can take the values $\mu$, $\nu$. 
In the case at hand, the frequency $\gamma_N$ of each oscillator, as well as the forcing terms $\xi_N$, depend on the slow time $T$ and the mode's wave number $k$. Explicitly, we have for the frequencies
\begin{subequations}
\begin{align}
    \gamma_{\mu}^2(k,T;\epsilon) \equiv \left(P_{11}(k,T;\epsilon)+\epsilon^2 \frac{\Lambda^{(-1)}_{11}(2-\Lambda^{(-1)}_{11})}{4T^2} \right)~,\\
     \gamma_{\nu}^2(k,T;\epsilon)\equiv\left(P_{22}(k,T;\epsilon)+\epsilon^2 \frac{\Lambda^{(-1)}_{22}(2-\Lambda^{(-1)}_{22})}{4T^2} \right)~,
\end{align}
\end{subequations}
%
%
%
%
%
%
%
%
%
%
%
while the forcing terms read
\begin{subequations}\label{Eq:forcing_terms_xiN}
\begin{align}   
&\epsilon\,\xi_\mu(k,T; \epsilon)\equiv - \epsilon\, (-T)^{\frac{1}{2}\left(\Lambda^{(-1)}_{11}-\Lambda^{(-1)}_{22}\right)}\left[ T^{-1}\Lambda^{(-1)}_{12}\dot{\nu}+\left(\frac{1}{\epsilon}P_{12}(k,T; \epsilon)-\epsilon\frac{\Lambda^{(-1)}_{22}\Lambda^{(-1)}_{12}}{2T^2} \right)\nu \right] ~,\\[5mm]
&\epsilon\,\xi_\nu(k,T; \epsilon) \equiv -\epsilon\, (-T)^{\frac{1}{2}\left(\Lambda^{(-1)}_{22}-\Lambda^{(-1)}_{11}\right)}\left[ T^{-1}\Lambda^{(-1)}_{21}\dot{\mu}+\left(\frac{1}{\epsilon}P_{21}(k,T; \epsilon)-\epsilon\frac{ \Lambda^{(-1)}_{11}\Lambda^{(-1)}_{21}}{2T^2}\right)\mu\right]   ~. 
\end{align}
\end{subequations}
%
%
%
%
Given the above expressions, we may proceed to compute multiple-scale solutions to the coupled ODE system if we further assume that the off-diagonal elements of the matrix $P_{ij}$ are $\mathcal{O(\epsilon)}$, such that the forcing terms $\xi_N$ in \eqref{Eq:forcing_terms_xiN}
do not contain negative powers of $\epsilon$. In this way, the overall magnitude of the $\xi_N$ is of order $\mathcal{O}(\epsilon)$~.

Next, we introduce the action and angle variables corresponding to the tensor modes $\mu$ and $\nu$, to bring our system into \textit{standard form} as outlined in Ref.~\cite{kevorkian_multiple_1996}:
\begin{subequations}\label{Eq:actionangle_defs}
\begin{alignat}{2}
    &\mathcal{J}_{\mu} \equiv \frac{\dot{\mu}^2+\gamma_{\mu}^2\mu^2}{2\gamma_{\mu}}~,\qquad
    &&\Psi_{\mu} \equiv \tan^{-1}\left(\frac{\gamma_{\mu}\mu}{\dot{\mu}}\right)~,\\
    &\mathcal{J}_{\nu} \equiv  \frac{\dot{\nu}^2+\gamma_{\nu}^2\nu^2}{2\gamma_{\nu}}~,\qquad
    &&\Psi_{\nu} \equiv  \tan^{-1}\left(\frac{\gamma_{\nu}\nu}{\dot{\nu}}\right)~.
\end{alignat}
\end{subequations}
In terms of these variables, 
we recast the system of two second-order equations~\eqref{Eq:SecondOrderEoM} as a system of four first-order equations:
%
%
\begin{subequations}\label{Eq:actionangle_system}
\begin{align}
    \dot{\mathcal{J}}_{\mu}&=-\epsilon\frac{\gamma^\prime _\mu}{\gamma_\mu}\mathcal{J}_\mu \cos(2\Psi_\mu)+\epsilon\,\xi_\mu\sqrt{\frac{2\mathcal{J}_\mu}{\gamma_\mu}} \cos(\Psi_\mu)~,\\
    \dot{\mathcal{J}}_{\nu}&=-\epsilon\frac{\gamma^\prime _\nu}{\gamma_\nu}\mathcal{J}_\nu \cos(2\Psi_\nu)+\epsilon\,\xi_\nu\sqrt{\frac{2\mathcal{J}_\nu}{\gamma_\nu}} \cos(\Psi_\nu)~,\\
   \dot{\Psi}_\mu&=\gamma_\mu +\epsilon \frac{\gamma^\prime _\mu}{2\gamma_\mu}\sin(2\Psi_\mu)-\epsilon\frac{\xi_\mu}{\sqrt{2 \gamma_\mu \mathcal{J}_\mu}}\sin(\Psi_\mu)~,\\
    \dot{\Psi}_\nu&=\gamma_\nu +\epsilon\frac{\gamma^\prime _\nu}{2\gamma_\nu}\sin(2\Psi_\nu)-\epsilon\frac{\xi_\nu}{\sqrt{2 \gamma_\nu \mathcal{J}_\nu}} \sin(\Psi_\nu)~.
\end{align}
\end{subequations}
Here and in the following, a prime is used as a short-hand notation for differentiation with respect to $T$, that is $\phantom{\cal J}^\prime\equiv\frac{\de}{\de T}$~. It is the action variables that are adiabatic invariants, since $\dot{\mathcal{J}}_{N}={\cal O}(\epsilon)$, and thus to leading order may only depend on the slow time $T$~. 
This permits us to treat the fast ($t$) and slow ($T$) times as independent variables at each order of $\epsilon$~.
The system \eqref{Eq:actionangle_system} belongs to the more general class of dynamical systems
\begin{subequations}\label{Eq:actionangle_system_general}
\begin{align}
    \dot{\mathcal{J}}_N&=\epsilon\, F_N(\mathcal{J}_i,\Psi_i,T;\epsilon)~,\\
    \dot{\Psi}_N&=\gamma^{(0)}_N+\epsilon\, G_N(\mathcal{J}_i,\Psi_i,T;\epsilon)~.
\end{align}
\end{subequations}
In our particular case, the functions $F_N$ and $G_N$ take the functional form 

\begin{subequations}
\begin{align}
    F_{\mu}&=-\frac{\gamma^\prime _\mu}{\gamma_\mu}\mathcal{J}_\mu \cos(2\Psi_\mu)+\xi_\mu\sqrt{\frac{2\mathcal{J}_\mu}{\gamma_\mu}} \cos(\Psi_\mu)~,\\
    F_{\nu}&=-\frac{\gamma^\prime _\nu}{\gamma_\nu}\mathcal{J}_\nu \cos(2\Psi_\nu)+\xi_\nu\sqrt{\frac{2\mathcal{J}_\nu}{\gamma_\nu}} \cos(\Psi_\nu)~,\\
   G_\mu&=\frac{\gamma_\mu-{\gamma^{(0)}_\mu}}{\epsilon}
   +  \frac{\gamma^\prime _\mu}{2\gamma_\mu}\sin(2\Psi_\mu)-\frac{\xi_\mu}{\sqrt{2\gamma_\mu \mathcal{J}_\mu}}\sin(\Psi_\mu) \label{Gmu}~,\\
    G_\nu&= \frac{\gamma_\nu-{\gamma^{(0)}_\nu}}{\epsilon}
    + \frac{\gamma^\prime _\nu}{2\gamma_\nu}\sin(2\Psi_\nu)- \frac{\xi_\nu}{\sqrt{2\gamma_\nu \mathcal{J}_\nu}}\sin(\Psi_\nu) \label{Gnu}~.
\end{align}
\end{subequations}
It is useful to note here that the first terms on the right hand side of Eqs.~\eqref{Gmu} and \eqref{Gnu} are regular in the small-$\epsilon$ limit, as we can easily check considering the perturbative $\epsilon$-expansion of each $\gamma_N$, which gives $\frac{\gamma_N - \gamma^{(0)}_N}{\epsilon}=\gamma^{(1)}_N + \epsilon\, \gamma^{(2)}_N + \mathcal{O}(\epsilon^2)$~. The unperturbed frequencies are given by ${\gamma^{(0)}_\mu}=\sqrt{{P^{(0)}_{11}}}$ and ${\gamma^{(0)}_\nu}=\sqrt{{P^{(0)}_{22}}}$~.

%

\subsection{Multiple-scale expansion of the dynamical system}\label{Sec:3dot2}
To ensure the applicability of multiple-scale analysis for systems of this kind, we shall require that $F_N$ and $G_N$ satisfy the following two conditions \cite{kevorkian_multiple_1996}:
\begin{itemize}
    \item[i)] $F_N$ and $G_N$ are $\mathcal{O}(1)$ in the limit $\epsilon \to 0$~;
    \item[ii)] $F_N$ and $G_N$ are periodic functions of $\Psi_i$ with period $2\pi$~.
\end{itemize}
The reason for such conditions can be understood as follows.
These functions must be $\mathcal{O}(1)$ in the limit $\epsilon \rightarrow 0$ to ensure that the perturbative expansion does not contain singular terms,
and also to ensure that the leading-order terms in \eqref{Eq:actionangle_system_general} are not of the same order as the perturbations. 
Periodicity with respect to each $\Psi_i$ ensures that each function $F_N$ and $G_N$ may be decomposed uniquely into an \textit{average} ($\bar{F}_N$, $\bar{G}_N$) and \textit{oscillating} ($\hat{F}_N$, $\hat{G}_N$) term. 
The oscillating components are defined such that they have zero average with respect to each angle variable $\Psi_i$~. For both $F_N$ and $G_N$ this decomposition is defined as
\begin{equation}\label{Eq:FNdecomposition_general}
    F_N(\mathcal{J}_i,\Psi_i,T;\epsilon) = \Bar{F}_N(\mathcal{J}_i,T;\epsilon)+\hat{F}_N(\mathcal{J}_i,\Psi_i,T;\epsilon)~,
\end{equation}
where 
\begin{equation}\label{Eq:FNaveraged_general}
    \Bar{F}_N(\mathcal{J}_i,T;\epsilon)\equiv \frac{1}{(2\pi)^2}\int_0^{2\pi}\de\Psi_\mu\int_0^{2\pi}\de\Psi_\nu\;F_N(\mathcal{J}_i,\Psi_i,T;\epsilon) ~.
\end{equation}
Evaluating explicitly the averaged and oscillating components of $F_N$ and $G_N$ for the class of models at hand, as given in Eq.~\eqref{Eq:actionangle_system_general}, we obtain
\begin{subequations}\label{Eq:actionangle_barhatcomponents}
\begin{align}
    &\Bar{F}_N(\mathcal{J}_i,T;\epsilon)=0~,\label{Eq:actionangle_barhatcomponents_a}\\
    &\Bar{G}_N(\mathcal{J}_i,T;\epsilon)=\frac{\gamma_N-{\gamma^{(0)}_N}}{\epsilon}~,\label{Eq:actionangle_barhatcomponents_b}\\
    &\hat{F}_N(\mathcal{J}_i,\Psi_i,T;\epsilon)=-\frac{\gamma^\prime _N}{\gamma_N}\mathcal{J}_N \cos(2\Psi_N)+\xi_N\sqrt{\frac{2\mathcal{J}_N}{\gamma_N}} \cos(\Psi_N)~,\\
    &\hat{G}_N(\mathcal{J}_i,\Psi_i,T;\epsilon)=\frac{\gamma^\prime _N}{2\gamma_N}\sin(2\Psi_N)-\frac{\xi_N}{\sqrt{2\gamma_N \mathcal{J}_N}}\sin(\Psi_N)~.
\end{align}
\end{subequations}
We note that the coupling between the two oscillators only enters the expression of the oscillating components $\hat{F}_N$, $\hat{G}_N$~.

The following perturbative ansatz is then assumed for the solutions $\mathcal{J}_N$ and $\Psi_N$, motivated by the independence of the slow and fast timescales, and the 
smallness of $\epsilon$,
\begin{subequations}\label{Eq:pertexp_action_angle}
\begin{align}
   &\mathcal{J}_N(t;\epsilon)={\mathcal{J}^{(0)}_N}(T)+\epsilon\, {\mathcal{J}^{(1)}_N}(\tau_i,T) +\epsilon^2\, {\mathcal{J}^{(2)}_N}(\tau_i,T) + \mathcal{O}(\epsilon^3)~,\\
   &\Psi_N(t;\epsilon)={\Psi^{(0)}_N}(\tau_N, T)+\epsilon\, {\Psi^{(1)}_N}(\tau_i,T) +\epsilon^2\, {\Psi^{(2)}_N}(\tau_i,T)+ \mathcal{O}(\epsilon^3)~.
\end{align}
\end{subequations}
Note that the zero-th order contribution to $\mathcal{J}_N$ depends only on $T$ as a consequence of its adiabatic invariance to leading order.\footnote{Adiabatic invariants are quantities whose derivative is purely oscillatory with zero average on the fast timescale \cite{kevorkian_multiple_1996}.} Here, the $\tau_i$ are the characteristic fast times of each oscillator, and are associated with the $\gamma_i$ frequencies. These can be formally defined as
\begin{equation}\label{Eq:Omega_definition}
    \frac{d\tau_N}{dt}=\Omega_N(T(t)) ~,
\end{equation}
where $\Omega_N$ are functions of the slow time coordinate, to be determined later. (Eventually, when Eq.~\eqref{Eq:Omega_definition} is solved, one should substitute $T=\epsilon\, t$.)
It is important to note that, while our approximations will only be evaluated up to linear order, the second-order terms in the expansion of the action and angle variables \eqref{Eq:pertexp_action_angle} are nonetheless necessary, as we must proceed to expand the full system of differential equations \eqref{Eq:actionangle_system_general} up to $\mathcal{O}(\epsilon^2)$ in order to fully determine the integration functions that arise from the first-order perturbative equations. This procedure will be outlined in greater clarity in the following subsections.

Similarly to Eq.~\eqref{Eq:pertexp_action_angle}, we shall also proceed to expand perturbatively both the averaged and oscillating components of $F_N$ and $G_N$ for the class of models at hand, as given in Eq.~\eqref{Eq:actionangle_barhatcomponents}. Equations~\eqref{Eq:actionangle_barhatcomponents_a}, \eqref{Eq:actionangle_barhatcomponents_b} imply for the averaged components
\begin{subequations}
\begin{align}
    \Bar{F}_N^{(n)}(\mathcal{J}_i,T;\epsilon)=0~,\\
    \bar{G}_N^{(n)}=\gamma_N^{(n+1)}~,
\end{align}
\end{subequations}
for all orders $n$. The perturbative expansion of the oscillating components $\hat{F}_N$ and $\hat{G}_N$ yields lengthy expressions. Their zeroth-order expansion coefficients are reported in Appendix~\ref{Sec:AppendixA}. 
The explicit expressions of higher-order terms are not needed to determine the solutions to the order of perturbation theory we are interested in, as explained below.


We may now define the total time derivative operator. Regarding $t$ as a function of the fast and slow times---regarded as independent---applying the chain rule and using Eq.~\eqref{Eq:Omega_definition}, we have
\begin{equation}\label{Eq:totaltimeder}
    \frac{\de}{\de t}=\Omega_\mu (T)\frac{\partial}{\partial \tau_\mu}+\Omega_\nu(T)\frac{\partial}{\partial \tau_\nu}+\epsilon\,\frac{\partial}{\partial T}~.
\end{equation}
This equation is central to 
the method of multiple scales, as it formally defines the different timescales the system evolves along and treats them as independent variables. 
A multiple-scale expansion is able to account for the cumulative effect of small perturbations and gives a uniformly valid approximation over long timescales $t \simeq \epsilon^{-1}$. The perturbations are treated as having an additional dependence on the slow time variable $T$, as well the 
fast time variables $\tau_i$
over the full, unperturbed system. This additional freedom is exploited to remove secular terms at each order in $\epsilon$ (which would otherwise grow monotonically with $t$).

Finally, substituting Eq.~\eqref{Eq:totaltimeder} into the system~\eqref{Eq:actionangle_system_general}, decomposing the equations into averaged and oscillating components and then expanding each term perturbatively, we collect terms of equal order in $\epsilon$ and return the following equations for each order in the perturbative expansion
\begin{subequations}
\begin{align}
    \mathcal{O}(1):\quad&\frac{\partial{\Psi^{(0)}_N}}{\partial \tau_N}\Omega_N = {\gamma^{(0)}_N}~,\label{O1}\\
    \mathcal{O}(\epsilon):\quad &\frac{\de{\mathcal{J}^{(0)}_N}}{\de T}+\Omega_\mu \frac{\partial {\mathcal{J}^{(1)}_N}}{\partial \tau_\mu} +\Omega_\nu \frac{\partial {\mathcal{J}^{(1)}_N}}{\partial \tau_\nu} = {\hat{F}^{(0)}_N}~,\label{pOe}\\
    &\frac{\partial {\Psi^{(0)}_N}}{\partial T}+\Omega_\mu \frac{\partial {\Psi^{(1)}_N}}{\partial \tau_\mu} +\Omega_\nu \frac{\partial {\Psi^{(1)}_N}}{\partial \tau_\nu} ={\Bar{G}^{(0)}_N}+{\hat{G}^{(0)}_N}~,\label{qOe}\\
   \mathcal{O}(\epsilon^2):  \quad &\frac{\partial {\mathcal{J}^{(1)}_N}}{\partial T}+\Omega_\mu \frac{\partial {\mathcal{J}^{(2)}_N}}{\partial \tau_\mu} +\Omega_\nu \frac{\partial {\mathcal{J}^{(2)}_N}}{\partial \tau_\nu} ={\hat{F}^{(1)}_N}~,\label{pOe2}\\
    &\frac{\partial {\Psi^{(1)}_N}}{\partial T}+\Omega_\mu \frac{\partial {\Psi^{(2)}_N}}{\partial \tau_\mu} +\Omega_\nu \frac{\partial {\Psi^{(2)}_N}}{\partial \tau_\nu} ={\Bar{G}^{(1)}_N}+{\hat{G}^{(1)}_N} ~.\label{qOe2}
\end{align}
\end{subequations}
To solve this system of PDEs, we must ensure that we avoid secular terms at each order of $\epsilon$ for each action and angle variable. This is done by removing terms that would otherwise grow unbounded with the fast times $\tau_N$~. This ensures that the approximation is uniformly valid over an interval $1/\epsilon$~.

\subsection{Solving the perturbative equations}
The solution to Eq.~\eqref{O1} is as follows
\begin{equation}
\label{O1Solution}
    {\Psi^{(0)}_N}=d_N(T)\tau_N+{f^{(0)}_{\Psi_N}}(T)~, \quad \text{where}\;\;d_N(T)=\frac{{\gamma_N}^{(0)}}{\Omega_N(T)}~.
\end{equation}
The integration function
${f^{(0)}_{\Psi_N}}(T)$ is undetermined at this stage, but can be determined explicitly by substituting the expression for  ${\Psi^{(0)}_N}$ into Eq.~\eqref{qOe}, which gives
\begin{equation}
\label{subbedeq}
    d^{\prime}_N(T)\tau_N+{f^{(0)\prime}_{\Psi_N}}(T) + \Omega_\mu \frac{\partial {\Psi^{(1)}_N}}{\partial \tau_\mu} +\Omega_\nu \frac{\partial {\Psi^{(1)}_N}}{\partial \tau_\nu} = {\Bar{G}^{(0)}_N}+{\hat{G}^{(0)}_N}~.
\end{equation}
The solution for ${\Psi^{(1)}_N}$ would involve quadratic terms in the $\tau_i$ unless we fix ${d^{\prime}_N}(T)=0$~. Therefore, we must remove such terms by requiring that $d^{\prime}_N$ is a constant; without loss of generality, we can set $d_N(T)=1$~. In turn, this allows to determine the frequencies $\Omega_N(T)$ using \eqref{O1Solution}
\begin{equation}
    \Omega_\mu={\gamma^{(0)}_\mu}=\sqrt{{P^{(0)}_{11}}}  ~, \quad  \Omega_\nu={\gamma^{(0)}_\nu}=\sqrt{{P^{(0)}_{22}}}~.
\end{equation}
Next, we take the averaged component of Eq.~\eqref{subbedeq} over the angle variables,
\begin{equation}
    {f^{(0)\prime}_{\Psi_N}}=\bar{G}_N^{(0)}.
\end{equation}
which is solved by quadrature
\begin{equation}
    {f^{(0)}_{\Psi_N}}(T)=c^{(0)}_N+\int_{T_0}^{T}\de u\;\bar{G}_N^{(0)}(u)~,
\end{equation}
where $c^{(0)}_N$ arises as an integration constant.
Proceeding in a similar fashion as above, we can show that the first term in Eq.~\eqref{pOe} must be vanishing, otherwise $\mathcal{J}^{(1)}_N$ would contain terms linear in the $\tau_i$.
%
In summary, the zeroth-order solutions read 
\begin{subequations}
    \begin{align}
        &{\mathcal{J}^{(0)}_N}={\zeta^{(0)}_N}~,\\
         &{\Psi^{(0)}_N}(\tau_N,T)=\tau_N+f^{(0)}_{\Psi_N}~,
    \end{align}
\end{subequations}
with $\tau_\mu=\int_{t_0}^{t}\de \tilde{t}\; \Omega_\mu (\epsilon\, \tilde{t})$ and $\tau_\nu=\int_{t_0}^{t}\de \tilde{t} \; \Omega_\nu (\epsilon\,\tilde{t})$~.


The next step in the computation of the solution for the angle variables at $\mathcal{O}(\epsilon)$ is to integrate the oscillatory component of Eq.~\eqref{subbedeq}. This leads to 
\begin{equation}\label{Eq:firstordersol_angle}
    {\Psi^{(1)}_N}(\tau_i, T)={f^{(1)}_{\Psi_N}}(T)+\int\de s\; \hat{G}^{(0)}_N(\mathcal{J}^{(0)}_i,\Omega_i s+f^{(0)}_{\Psi_i},T)\Big\lvert_{ \Omega_i s= \tau_i}~,
\end{equation}
where ${f^{(1)}_{\Psi_N}}(T)$ is to be determined by consistency conditions at $\mathcal{O}(\epsilon^2)$~.
%
%
Similarly, we solve the oscillatory component of Eq.~\eqref{pOe} and express the solution for the action variables in the form
\begin{equation}\label{Eq:firstordersol_action}
    {\mathcal{J}^{(1)}_N}(\tau_i,T)={f^{(1)}_{\mathcal{J}_N}}(T)+\int\de s\;\hat{F}^{(0)}_N(\mathcal{J}^{(0)}_i,\Omega_i s+f^{(0)}_{\Psi_i},T)\Big\lvert_{\Omega_i s=\tau_i}~,
\end{equation}
where ${f^{(1)}_{\mathcal{J}_N}}(T)$ are as yet undetermined functions, which will be determined from the second order equations.
We can explicitly evaluate the integrals on $\hat{F}^{(0)}_N$ and $\hat{G}^{(0)}_N$ that appear in Eqs.~\eqref{Eq:firstordersol_angle}, \eqref{Eq:firstordersol_action}. The variable that we integrate over is determined parametrically by the equation $\Omega_i s = \tau_i$, which describes a line in the $(\tau_{\mu},\tau_{\nu})$ plane. The integrations can be carried out analytically, and the results are reported in Appendix~\ref{Sec:AppendixA}. 

We can now complete the derivation of the first-order solutions, using the second-order equations to determine the remaining free integration functions. We extract the oscillatory component of Eq.~\eqref{pOe2} and set it to zero, in order to ensure that secular terms are removed from ${\mathcal J}_N^{(2)}$. This gives $\frac{\partial {\mathcal{J}^{(1)}_N}}{\partial T}=0$, which, using Eq.~\eqref{Eq:firstordersol_action}, implies
%
%
%
\begin{equation}
    {f^{(1)\, \prime}_{\mathcal{J}_N}}=0~.
\end{equation}
Hence,
\begin{equation}
     {f^{(1)}_{\mathcal{J}_N}}=\zeta_N^{(1)}~,
\end{equation}
where $\zeta_N^{(1)}$ are integration constants. Similarly, to ensure the absence of secular terms in $\Psi_N^{(2)}$, we extract the averaged components of both sides of Eq.~\eqref{qOe2} and set them equal, which gives the following condition $\frac{\partial {\Psi^{(1)}_N}}{\partial T}={\Bar{G}^{(1)}_N}$~. Substituting Eq.~\eqref{Eq:firstordersol_angle}, we obtain an equation to determine the remaining integration function
\begin{equation}
    {f^{(1)\, \prime}_{\Psi_N}}={\bar{G}_N}^{(1)}~,
\end{equation}
whose solution is obtained by quadrature,
\begin{equation}
     {f^{(1)}_{\Psi_N}}(T)=c^{(1)}_N+\int_{T_0}^{T}\de u\;{\bar{G}_N}^{(1)}(u)~,
\end{equation}
where $c^{(1)}_N$ arises as an integration constant.
This concludes the derivation of the full solution to first order in $\epsilon$.

\subsection{Results summary and discussion}
We have obtained general approximate solutions to first order in $\epsilon$, for a system of two coupled helicity-2 modes within the class of models described by Eq.~\eqref{Eq:SecondOrderEoM}, while remaining agnostic to any particular choice of a model within this class. The general solutions for the action and angle variables read
\begin{subequations}\label{Eq:fullsolutions_JPsi}
    \begin{align}
        &{\mathcal{J}_N}(\tau_i,T;\epsilon)= {\zeta^{(0)}_N}+\epsilon\left({\zeta^{(1)}_N} + \int \de s\; \hat{F}^{(0)}_N(\mathcal{J}^{(0)}_i, \Omega_i s + f^{(0)}_{\Psi_i},T)\Big\lvert_{\Omega_i s=\tau_i}\; \right) +\mathcal{O}(\epsilon^2)~,\\
    \begin{split}
        &{\Psi_N}(\tau_i, T;\epsilon)=c^{(0)}_N+\tau_N+\int_{T_0}^{T}\de u\; \bar{G}^{(0)}_N(u)\; \\
        &+\; \epsilon\left(c^{(1)}_N+\int_{T_0}^{T}\de u\;{\bar{G}_N}^{(1)}(u)+\int \de s\;{\hat{G}_N}^{(0)}(\mathcal{J}^{(0)}_i, \Omega_i s + f^{(0)}_{\Psi_i},T)\Big\lvert_{\Omega_i s=\tau_i}\;\right)+\mathcal{O}(\epsilon^2)~.
    \end{split}
    \end{align}
\end{subequations} 
Choosing a specific model within the class of models at hand, will in turn determine the functional form of the functions $G_N$ and $F_N$ in \eqref{Eq:actionangle_system_general}. We can then obtain explicit formulae for the evolution of GWs using the full analytical expressions for the integrals in Eq.~\eqref{Eq:fullsolutions_JPsi}, which are given in Eq.~\eqref{Eq:mixing_terms}.

We stress that these approximations for the action and angle variables are only valid provided the diagonal elements of the matrix $P_{ij}$ are $\mathcal{O}(1)$ to leading order; otherwise, the approximate solutions for $\mathcal{J}_N$ and $\Psi_N$ would contain singular terms in the $\epsilon\to0$ limit, inherited from the functional dependencies of $F^{(0)}_N$ and $G^{(0)}_N$ on each $\gamma^{(0)}_N$.\footnote{This does not represent a significant restriction for our purposes, since the hypothetical case where $P_{ij}$ has an ${\cal O}(\epsilon)$ diagonal element corresponds to a free point particle coupled to an oscillator (as opposed to two coupled oscillators), and therefore cannot describe the Fourier components of a field in a consistent perturbative scheme.}
Similarly, the off-diagonal elements of the matrix $P_{ij}$ must be at least $\mathcal{O}(\epsilon)$ to ensure the perturbative expansion is well balanced; otherwise, the leading-order terms would be of the same order as the perturbations and the approximation scheme would not be directly applicable.

Finally, solutions for the original tensor perturbations $h$, $s$ can be obtained inverting Eqs.~\eqref{Eq:actionangle_defs} and \eqref{Eq:changevariables_munu}, which gives
\begin{subequations}\label{Inverse transformations to h and s}
\begin{align}
h&=\exp\left(-\int\de t\; \frac{\Lambda_{11}}{2k} \right) \lf(\frac{2{\cal J}_\mu}{\gamma_\mu}\rg)^{1/2}\sin(\Psi_\mu)~,\label{Eq:hGWsignal}\\
s&=\exp\left(-\int\de t\; \frac{\Lambda_{22}}{2k} \right) \lf(\frac{2{\cal J}_\nu}{\gamma_\nu}\rg)^{1/2}\sin(\Psi_\nu)~,
\end{align}
\end{subequations}
with ${\cal J}_N$ and $\Psi_N$ given by Eqs.~\eqref{Eq:fullsolutions_JPsi}. The time-evolution of the amplitude and phase of the GW signal (here represented by the $h$ mode) encode deviations from GR. We note that the solutions \eqref{Inverse transformations to h and s} are more general than those obtained in models with a single graviton \cite{Nishizawa:2017nef}, including scalar-tensor theories \cite{Arai:2017hxj} and parity-violating models \cite{Jenks:2023pmk}, where modulations of the phase and amplitude can be expressed as integrals of functions whose time-dependence is fixed by the evolution of the background alone. On the contrary, here the two tensor modes influence each other's evolution at each order in perturbation theory, as can be seen by the mixing terms contained in \eqref{Eq:mixing_terms}, and this is reflected in the time-dependence of the modes. Moreover, due to the presence of the extra dynamical field $s$, in general the evolution of $h$ depends on four initial conditions, instead of two as in single-graviton models.\footnote{However, in the case of bimetric gravity it is often assumed that perturbations of the metric uncoupled to matter fields are initially not excited at the emission of GWs \cite{Max:2017flc}.}
With these solutions at hand, one may proceed to explore the phenomenology of this class of models in greater detail, which will be the subject of future work. 
In the following Section, we demonstrate the validity of the approximate analytical solutions on some toy models by direct comparison with the numerical solutions.

\section{Applications to Phenomenological Models}\label{Sec:pheno_model}

We now turn our attention to the modifications to the phase and amplitude of the waveform of the tensorial perturbations, in specific examples which generalize some of the phenomenological models considered in Ref.~\cite{Ezquiaga:2021ler}. We consider cases of mass mixing (which is present, for instance, in Hassan-Rosen bimetric gravity), chiral mixing, velocity mixing and friction mixing, to demonstrate the accuracy of the approximations by comparing them with numerical results, and the regimes where the approximations break down. In all simulations in this section, we set the following initial conditions: $\mathcal{J}_\mu(t_0)=1$, $\mathcal{J}_\nu(t_0)=1/2$, $\Psi_\mu(t_0)=1/10$, $\Psi_\nu(t_0)=-1/20$~.

%

\subsection{Mass Mixing}

\begin{figure}
    \centering
    \includegraphics[width=1\linewidth]{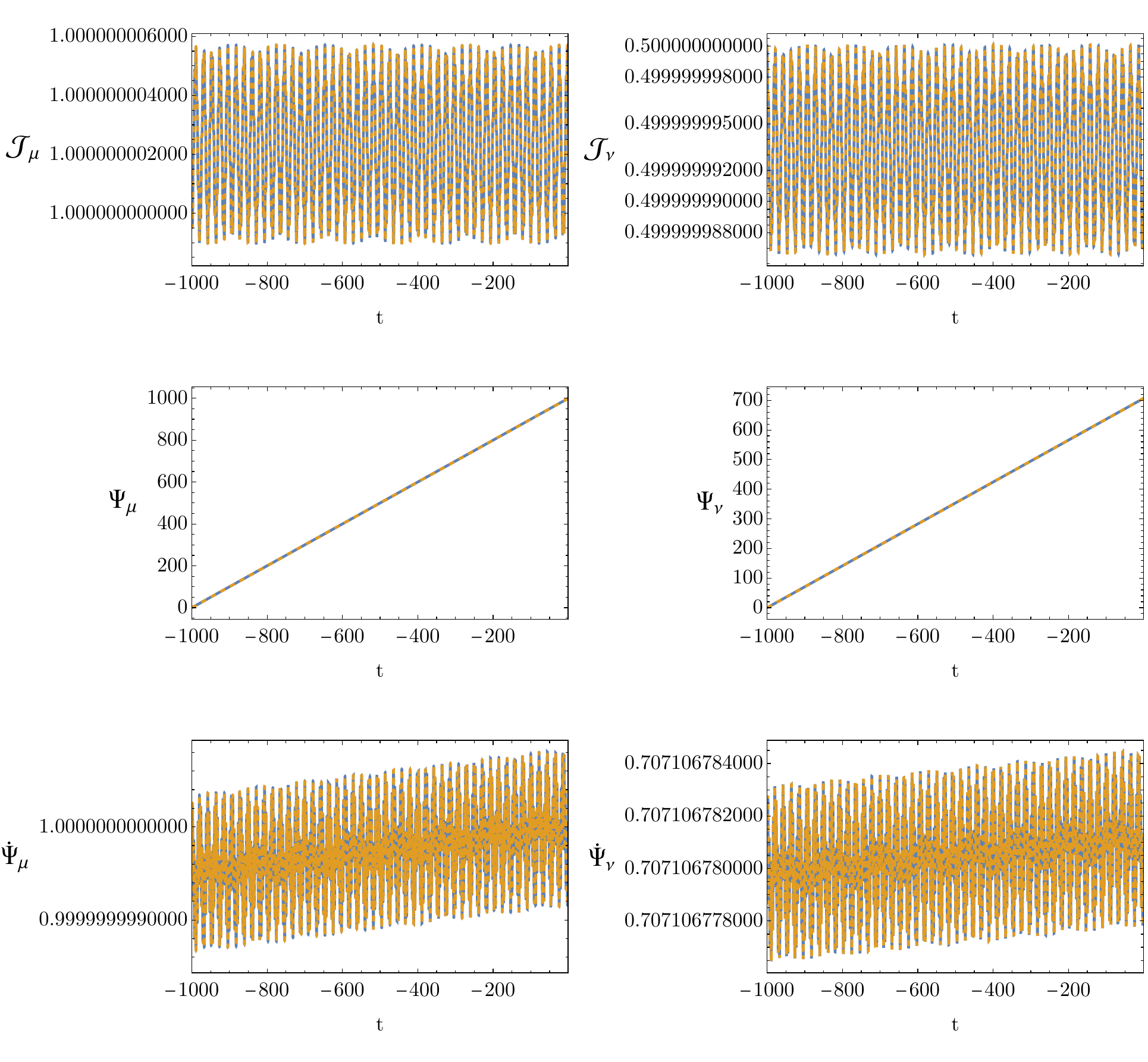}
    \caption{The numerical and approximate action, angle and $t$-derivative angle variables of Eq. \eqref{monomial time dependent model}, evolving along the dimensionless time parameter $t$.
     We have chosen the parameters $k=10^3$, $\epsilon=10^{-3}$, $t_0=-10^{3}$ and $m=1$~. The thick, blue curves correspond to the approximate solutions, and the dashed, orange curves correspond to the numerical solutions.}
    \label{Mass Mixing Monomial time Dependence Late Time}
\end{figure}

We consider a simple model of mass mixing with linear dependence on the slow time $T$, described by the toy model,
\begin{equation} \label{monomial time dependent model}
     \hat{\Lambda} = 0~, \quad
    \hat{\Pi} = 0~,\quad \hat{C}= \begin{pmatrix}
        1 & 0 \\
        0 & \frac{1}{2}
    \end{pmatrix} ~, \quad
    \hat{M} = \epsilon\, m^2 \begin{pmatrix}
       T & 1 \\
        2 & 2T
    \end{pmatrix}
   ~,
\end{equation} 
where the exotic perturbation $s$ propagates at a sub-luminal rate. The multiple-scale solutions for this model are given in Eq.~\eqref{Monomial Action and Angle}, and are plotted against the numerical solutions in Fig.~\ref{Mass Mixing Monomial time Dependence Late Time}. To solve the equations of motion numerically, we integrate from time $t_0=-10^3$ to the point of horizon crossing, $t=-1$~.
We plot the residuals for this model (that is, the absolute difference between the numerical and multiple-scale analytical solutions) in Fig.~\ref{Monomial Residuals} in Appendix~\ref{Sec:AppendixResidualPlots}. It is shown that the residuals are well below the scale of $\epsilon^2$ and therefore the evolution of the system is well-approximated to first order in the multiple-scale expansion. The corresponding solution for the GW signal $h$ is
\begin{equation}
    h(\tau_i,T;\epsilon) = \left(\frac{2\mathcal{J}_\mu}{\sqrt{P_{11}}}\right)^\frac{1}{2}\sin\left(\Psi_\mu\right)~.
    \label{GW signal no friction}
\end{equation}
We note that the explicit expression for the GW signal contains non-trivial modifications of the waveform and phase relative to GR. The reciprocal of the amplitude of this signal is proportional to the GW luminosity distance.
However, the full solution contains more information than the GW luminosity distance alone.
Moreover, in this class of models the GW luminosity distance is in general a function of the wavenumber (through the $P_{ij}$ matrix elements).
In this framework, the signal will necessarily include contributions from the velocity, chiral and mass matrix of the model studied.

\begin{figure}
    \centering
    \includegraphics[width=1\linewidth]{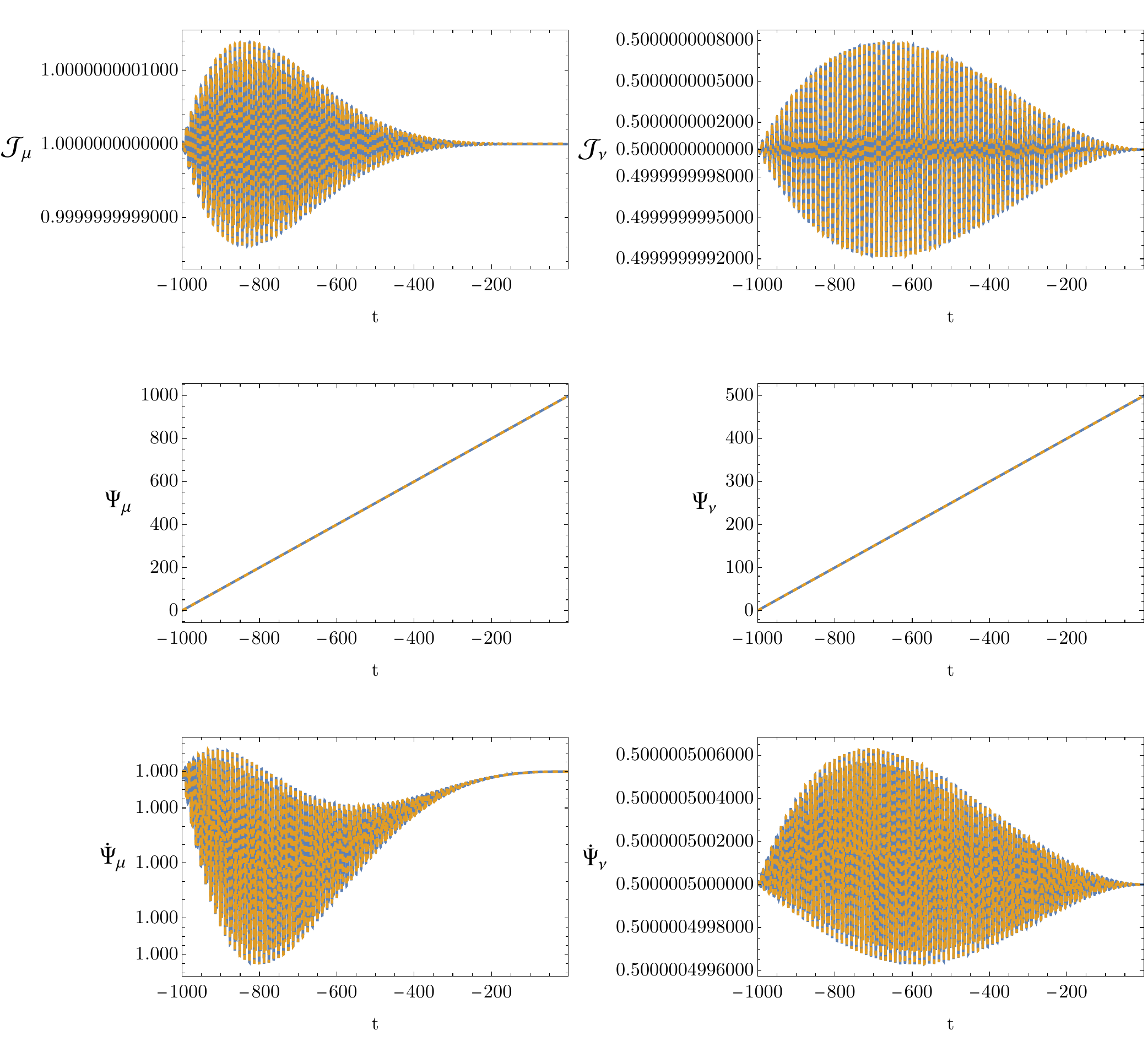}
     \caption{The numerical and approximate action, angle and $t$-derivative angle variables of Eq. \eqref{polynomial model}, evolving along the dimensionless time parameter $t$. For these plots we have selected the values $k=10^3$, $\epsilon=10^{-3}$, $m=1$, $t_0=-10^3$~. The thick, blue curves correspond to the approximate solutions, and the dashed, orange curves correspond to the numerical solutions.}
    \label{Time Dependent Model}
\end{figure}

We may take a less trivial example of mass mixing, where the mass matrix depends polynomially on the slow time, and
constant on-diagonal velocity and chiral terms are also included. One such example is
\begin{equation}\label{polynomial model}
    \hat{\Lambda} = 0 ~, \quad
    \hat{C}= \begin{pmatrix}
        1 & 0 \\
        0 & \frac{1}{4} 
    \end{pmatrix} ~, \quad
    \hat{\Pi} = \begin{pmatrix}
        \epsilon & 0 \\
        0 & \frac{\epsilon}{2}
    \end{pmatrix} ~, \quad
    \hat{M} = \epsilon\, m^2 \begin{pmatrix}
        T^3 + T^4 & T^5 + T^6 \\
        2(T^2 + T^3) & 2(T^4 + T^5)
    \end{pmatrix}~.
\end{equation} 
The approximate solutions to the model \eqref{polynomial model} are given explicitly in Eq.~\eqref{Polynomial Action and Angle}. The action and angle variables, and the $t$-derivatives of the angle variables, are plotted in Fig.~\ref{Time Dependent Model}, where the approximations are shown to match the numerical results well. The residuals are plotted in Fig.~\ref{Polynomial residuals} in Appendix~\ref{Sec:AppendixResidualPlots} and are shown to be below the scale of $\epsilon^2$. The GW signal for this model $h$ takes a similar form as Eq.~\eqref{GW signal no friction}, though with different functional dependencies on the characteristic times $\tau_i$ and the slow time $T$, which can be obtained combining Eq.~\eqref{Eq:hGWsignal} and Eq.~\eqref{Polynomial Action and Angle}.


\subsection{Chiral Mixing}
We proceed to test our results also on a simple toy model with chiral mixing. Consider the toy model
\begin{equation}\label{Chiral Model}
         \hat{\Lambda} = 
    \hat{M} = 0~, \quad 
    \hat{C}= \begin{pmatrix}
        1 & 0 \\
        0 & \frac{1}{2}
    \end{pmatrix}~,\quad
    \hat{\Pi} = \frac{ \epsilon}{T} \begin{pmatrix}
       \frac{3}{2} & 1 \\
        1 & \frac{1}{2}
    \end{pmatrix}~.
\end{equation}
The model in Eq.~\eqref{Chiral Model} is inspired by an analog model studied in Ref.~\cite{Ezquiaga:2021ler}.
\begin{figure}[t!]
\centering
\includegraphics[width=1\linewidth]{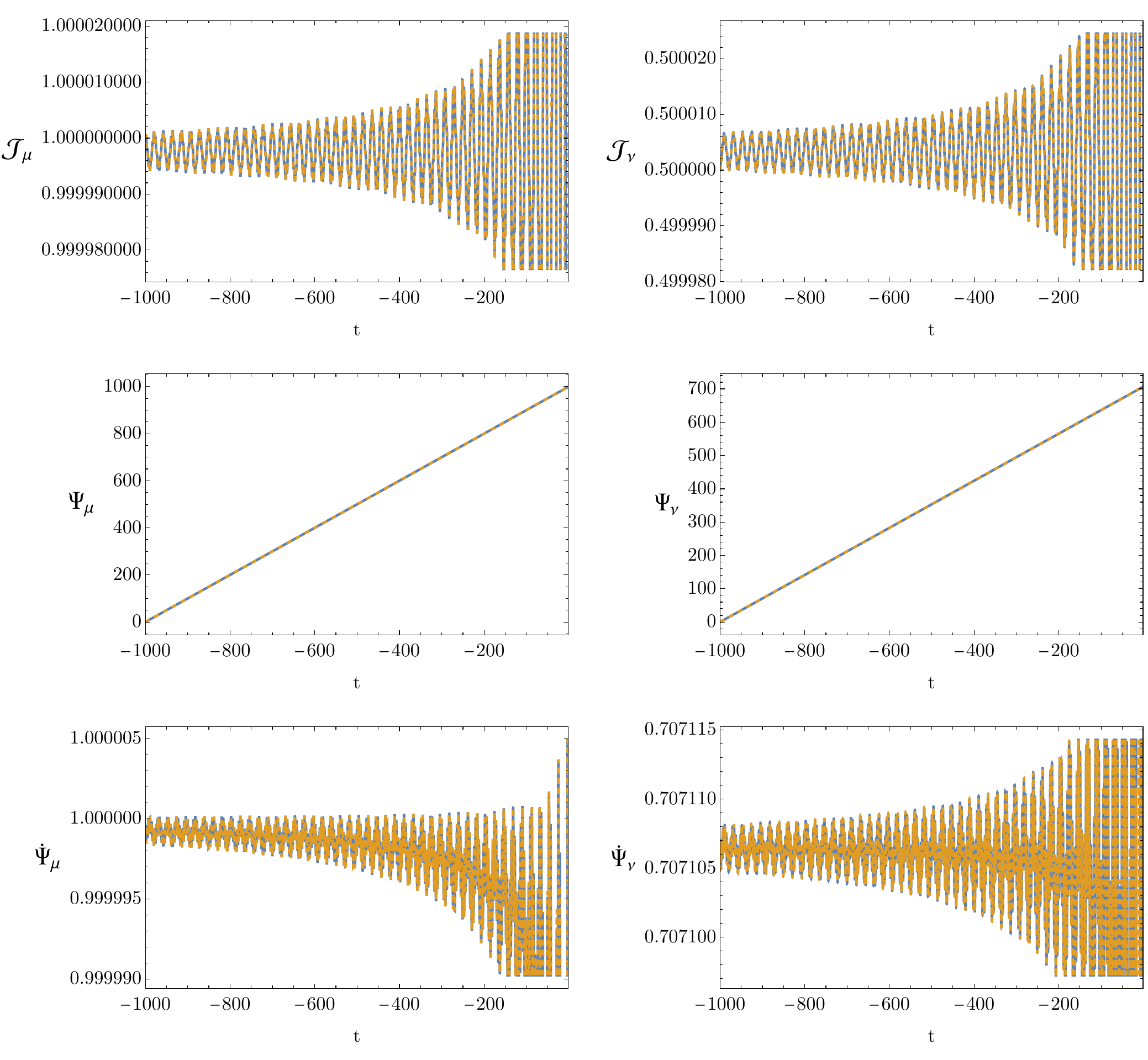}
    \caption{The numerical and approximate action, angle and $t$-derivative angle variables for the model in Eq.~\eqref{Chiral Model}, evolving along the dimensionless time parameter $t$. We have chosen $k=10^3$, $\epsilon=10^{-3}$, $t_0=-10^3$. The thick, blue curves correspond to the approximate solutions, and the dashed, orange curves correspond to the numerical solutions.}
    \label{Chiral Mixing Toy Model}
\end{figure}
The multiple-scale solutions for this model are given explicitly in Eq.~\eqref{Chiral Action and Angle}. We note that, in this case, the approximation retains the desired accuracy for a shorter time interval compared to the mass mixing model, and eventually fails before the modes exit the horizon. This is demonstrated by the residuals, which are plotted in Appendix~\ref{Sec:AppendixResidualPlots} in Fig.~\ref{Chiral Residuals}, where it is observed that in the late-time propagation history of a given mode, the residuals cross the threshold of $\epsilon^2$. This growth in the residuals obeys a power law.  The GW signal for this model $h$ takes a similar form as Eq.~\eqref{GW signal no friction}, with different functional dependencies on the characteristic fast times $\tau_i$ and the slow time $T$. Plotting these approximations against numerical results in Fig.~\ref{Chiral Mixing Toy Model}, we restrict to the region where the approximate solutions are valid to the desired accuracy, and thus cut off the plots at $t=-10^2$~.

\subsection{Velocity Mixing}
\begin{figure}
    \centering
    \includegraphics[width=1\linewidth]{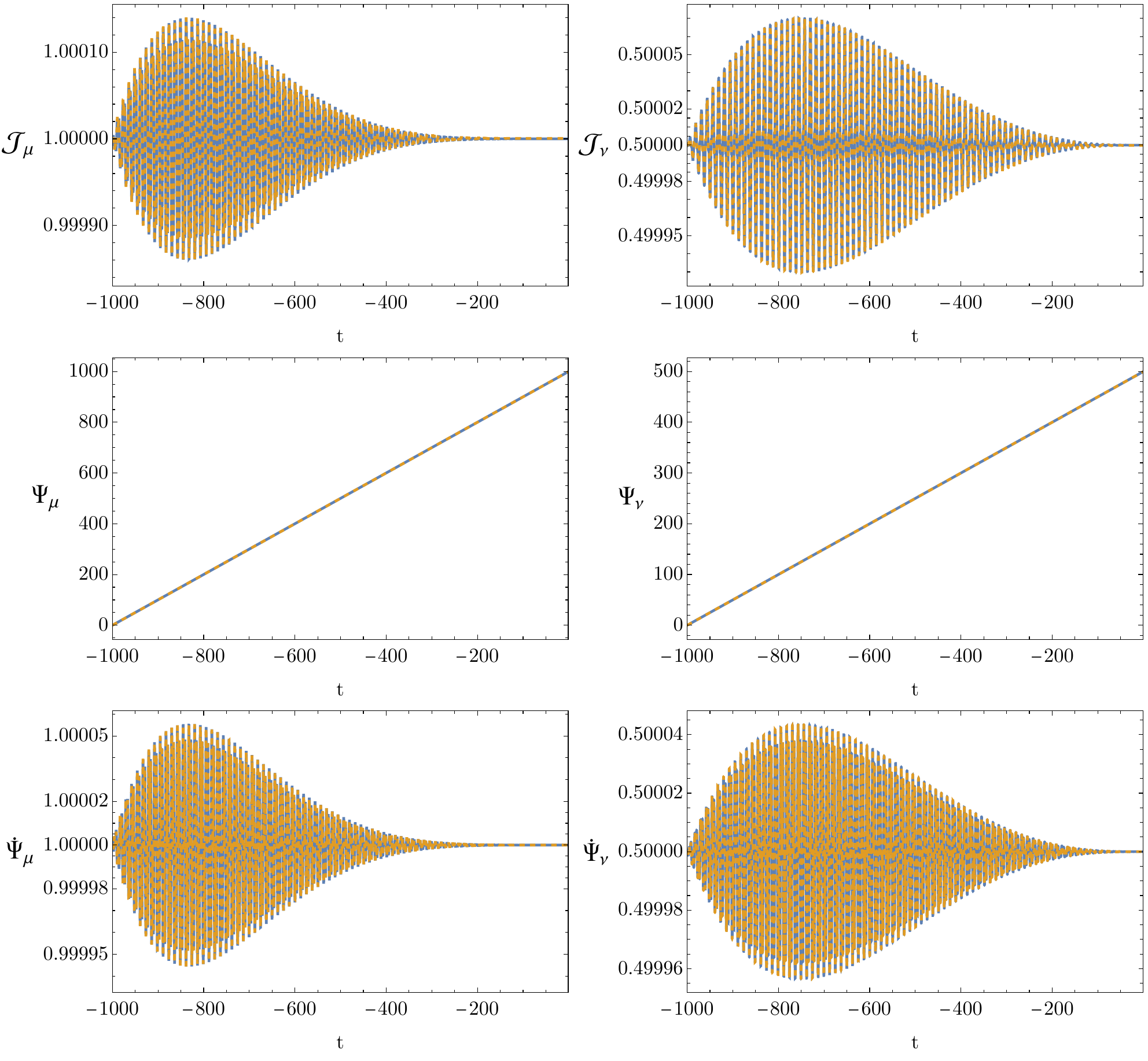}
    \caption{The numerical and approximate action angle and $t$-derivative angle variables of Eq. \eqref{velocity mixing model}, evolving along the dimensionless time parameter $t$. For these plots we have chosen the parameters $k=10^3$, $\epsilon=10^{-3}$, $t_0=-10^3$~. The thick, blue curves correspond to the approximate solutions, and the dashed, orange curves correspond to the numerical solutions.}
    \label{velocity mixing plots}
\end{figure}

We consider a toy model of velocity mixing with a polynomial time dependence on the off-diagonal entries,
\begin{equation}\label{velocity mixing model}
    \hat{\Lambda}=\hat{M}=\hat{\Pi}=0~, \quad 
    \hat{C}=\begin{pmatrix}
        1 &  \epsilon\, (T^5+T^6) \\
        \frac{1}{4}  \epsilon\, (T^3+T^4) &\frac{1}{4}
    \end{pmatrix} ~.
\end{equation}
These time-dependent terms are of order $\epsilon$, as this is a required assumption of our method, as discussed in earlier sections. The approximate solutions for this model are given by Eq.~\eqref{Velocity Action and Angle}. We plot the numerical and approximate solutions for the action and angle variables, as well as the $t$-derivative of angle variables in Fig.~\ref{velocity mixing plots}. 
The residuals of these solutions for the action and angle variables are plotted in Fig.~\ref{Velocity Residuals} in Appendix~\ref{Sec:AppendixResidualPlots}. The residuals are shown to oscillate around the scale of $\epsilon^2$, particularly in the angle variables. However, these residuals remain very close to the scale of $\epsilon^2$ throughout the time interval of interest, and thus satisfy our desired tolerance of the approximations. Also in this model, the GW signal $h$ takes a similar form as in Eq.~\eqref{GW signal no friction}, although with different functional dependencies on the characteristic times $\tau_i$ and the slow time $T$.

\subsection{Friction mixing}
\begin{figure}[t!]
\centering
    \includegraphics[width=1\linewidth]{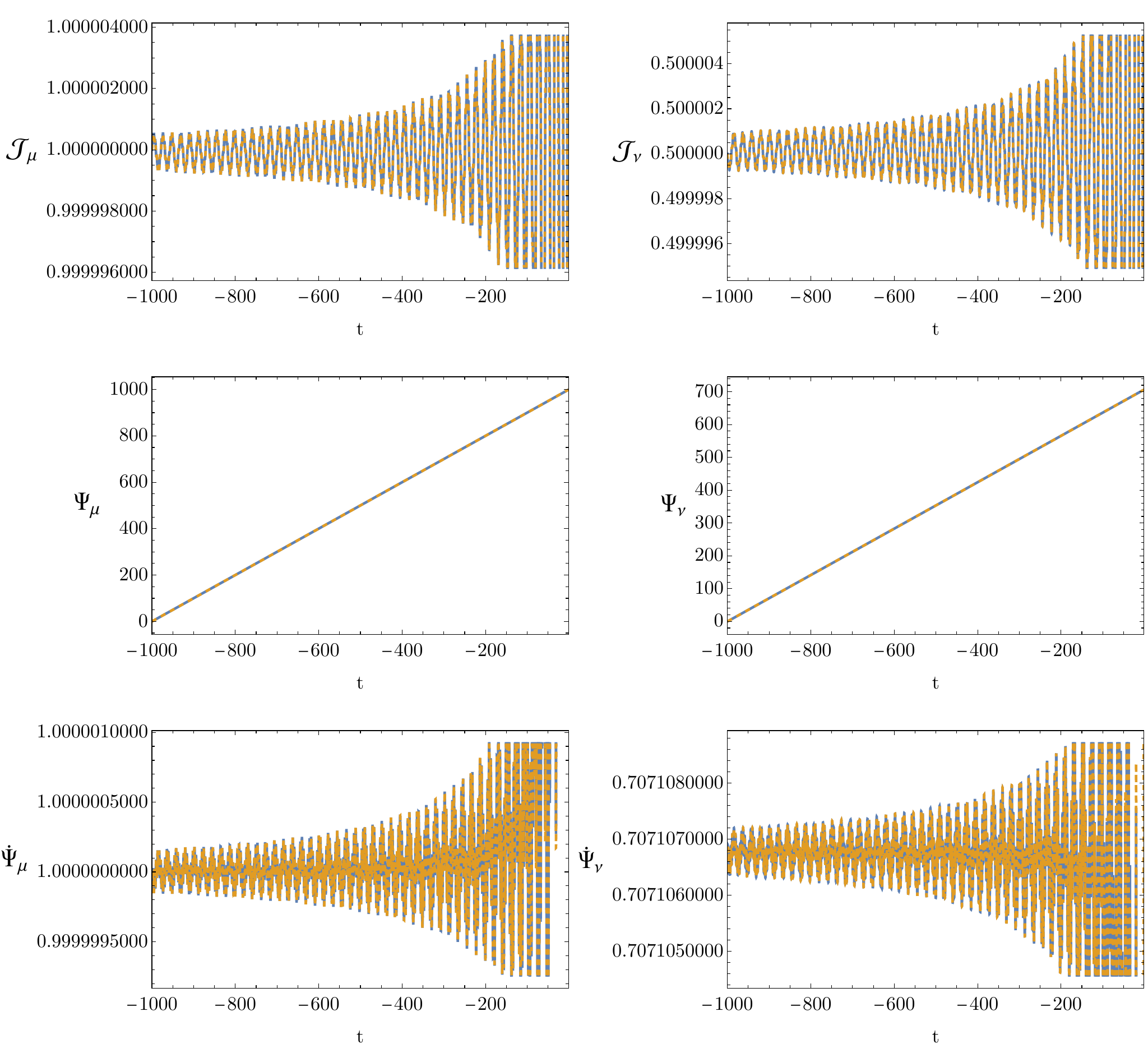}
    \caption{The action, angle and $t$-derivative angle variables of Eq. \eqref{Friction mixing model}, evolving along the dimensionless time parameter $t$. For these simulations we have chosen $k=10^3$, $\epsilon=10^{-3}$, $t_0=-10^3$, $\alpha=10^{-4}$ and $\Delta=10^{-2}$. The thick, blue curves correspond to the approximate solutions, and the dashed, orange curves correspond to the numerical solutions.}
    \label{Friction mixing plots}
\end{figure}

We consider the following toy model of friction mixing,
\begin{equation}\label{Friction mixing model}
    \hat{\Lambda}=\frac{1}{\eta}\begin{pmatrix}
        2(1-\Delta) & ~~ \frac{1}{4} \alpha \\ -\frac{1}{4} \alpha & 2(1+\Delta)
    \end{pmatrix}~, \quad
    \hat{C}=\begin{pmatrix}
        1 & 0 \\
        0 & \frac{1}{2}
    \end{pmatrix}~, \quad
    \hat{\Pi}=\hat{M}=0~.
\end{equation}
The first entry of the friction matrix, with $\Lambda^{(-1)}_{11}= 2(1-\Delta)$~, is written as such to recover the standard friction term of the GR, with a small deviation parameterized by $\Delta$~. The off-diagonal elements of the friction matrix are inspired by a model studied in Ref.~\cite{Ezquiaga:2021ler}, and $\alpha$ is a dimensionless parameter. The numerical and approximate solutions for the action and angle variables, as well as the $t$-derivative of the angle variables, are plotted in Fig.~\ref{Friction mixing plots}, where the plots are cut off at $t=-10^2$. The residuals are plotted in Fig.~\ref{Friction Residuals}. It is evident from the residuals that the action and angle variables are well-approximated throughout most of the propagation history of the mode until the modes approach horizon exit. However, the residuals in the action and angle variables exhibit a power-law growth, and cross the scale of $\epsilon^2$ while still in the sub-horizon regime. The reason for this is evident from the analytic expressions of the approximate solutions given in the Appendix in Eq.~\eqref{Friction Action and Angle}, and are inherited from the particular choice of model studied. Due to the particular choice made for the diagonal elements in the friction matrix, during the propagation history of the modes, the action and angle variables in both the $\mu$ and $\nu$ sectors are subject to a power-law decay which becomes dominant in the late time limit. Lastly, we observe that the GW signal $h$ for this model takes the form
\begin{equation}
    h(\tau_i,T;\epsilon) =  t^{\frac{\Delta - 2}{2k}}
   \left(\frac{2\mathcal{J}_\mu}{\sqrt{P_{11}}}\right)^\frac{1}{2}\sin\left(\Psi_\mu\right)~.
\end{equation}
In this case, the inclusion of frictional terms contributes to an extra power-law factor modifying the amplitude of the GW signal relative to previous models, which have no entries in the friction matrix.

\section{Conclusions}\label{Conclusions}

In this work, we have developed an approximation scheme to calculate the propagation of tensor perturbations on sub-horizon scales, in a class of models that exhibit an additional spin-2 partner to the standard massless spin-2 mode that propagates in GR, generalizing previous work \cite{Brizuela:2023uwt} in the context of bimetric gravity. 
We have assumed a universe dominated by dark energy, here described for simplicity as a positive cosmological constant.
This scheme was developed using a multiple-scale expansion, which employs two time variables $t$ and $T$, treated as independent. This allows us to determine approximations that are uniformly valid over the propagation history of the GW signal for long timescales, from emission to horizon exit. The method of multiple scales is particularly well-suited to describe the cumulative effect of small perturbations of coupled oscillator systems over long timescales (which, for sub-horizon modes, correspond to a large number of oscillations), where standard perturbative techniques as well as numerical simulations are bound to fail.

Our results show that, for the class of models at hand and unlike other examples of modified gravity theories, parameterizing the cosmological propagation of GWs in terms of a modified GW luminosity distance function alone may not be sufficient. As we have shown, more information is encoded in the action-angle variables that cannot be captured by the GW luminosity distance alone. 
As an example of possible applications of our methods, we considered some toy models with mass mixing, chiral mixing, velocity mixing, and friction mixing, which represent a generalization of the phenomenological models previously studied in Ref.~\cite{Ezquiaga:2021ler}. Direct comparison with numerical simulations of the system proved our analytical results to be accurate until horizon exit, except in the cases of chiral mixing and friction mixing. In these cases the approximations fail at a time before horizon exit.  

The solutions derived in this work may be specialized to concrete models and used to re-evaluate projected constraints on GW propagation in theories with interacting tensor fields, which have been derived earlier in Ref.~\cite{Ezquiaga:2021ler}. In addition, the evolution of the system in position space may reveal whether different components of $h$ contribute coherently to the GW signal, or can be temporally resolved at the detector, similarly to the analysis performed for Hassan-Rosen bimetric gravity in Ref.~\cite{Brizuela:2025cmz}.


\section*{Acknowledgements}
MdC acknowledges support from INFN iniziativa specifica GeoSymQFT. BS acknowledges support from a Science and Technologies Facilities Council (STFC) Doctoral Training Grant.  MS acknowledges support from the
Science and Technology Facility Council (STFC), UK, under
the research grant ST/X000753/1. 
\appendix

\renewcommand{\theequation}{\thesection.\arabic{equation}}

\section{Explicit expressions for \texorpdfstring{$\hat{F}^{(0)}_N$ and $\hat{G}^{(0)}_N$}{the oscillating components of F_N, G_N} and their line integrals}\label{Sec:AppendixA}
The functions $F_N$ and $G_N$ that appear as source terms in the first-order system~\eqref{Eq:actionangle_system_general} are decomposed into oscillating and averaged components as in Eq.~\eqref{Eq:FNdecomposition_general}. The averaged components and their perturbative expansions are given in Section~\ref{Sec:3dot2}.
The oscillating components are also expanded perturbatively in $\epsilon$ as follows
\begin{subequations}
\begin{align}
    \begin{split}
        \hat{F}_N(\mathcal{J}_i,\Psi_i,T;\epsilon)
        &={\hat{F}_N}^{(0)}(\mathcal{J}^{(0)}_i,\Psi^{(0)}_i,T;\epsilon) +\epsilon\,{\hat{F}_N}^{(1)}(\mathcal{J}^{(0)}_i,\mathcal{J}^{(1)}_i,\Psi^{(0)}_i,\Psi^{(1)}_i,T;\epsilon) \\
        &\phantom{=}  +\epsilon^2\,{\hat{F}_N}^{(2)}(\mathcal{J}^{(0)}_i,\mathcal{J}^{(1)}_i,\mathcal{J}^{(2)}_i,\Psi^{(0)}_i,\Psi^{(1)}_i,\Psi^{(2)}_i,T;\epsilon) + \mathcal{O}(\epsilon^3)~,
    \end{split}\\
    \begin{split}
        \hat{G}_N(\mathcal{J}_i,\Psi_i,T;\epsilon)
        &={\hat{G}_N}^{(0)}(\mathcal{J}^{(0)}_i,\Psi^{(0)}_i,T;\epsilon) +\epsilon\,{\hat{G}_N}^{(1)}(\mathcal{J}^{(0)}_i,\mathcal{J}^{(1)}_i,\Psi^{(0)}_i,\Psi^{(1)}_i,T;\epsilon) \\ 
        &\phantom{=} +\epsilon^2\,{\hat{G}_N}^{(2)}(\mathcal{J}^{(0)}_i,\mathcal{J}^{(1)}_i,\mathcal{J}^{(2)}_i,\Psi^{(0)}_i,\Psi^{(1)}_i,\Psi^{(2)}_i,T;\epsilon) + \mathcal{O}(\epsilon^3)~.
    \end{split}
\end{align}
\end{subequations}
To perform this expansion, it is useful to express also the forcing terms $\xi_N$ as functions of the action and angle variables\\

\begin{equation}
    \begin{split}
    \epsilon\,\xi_\mu(k,T; \epsilon)= - \epsilon\, (-T)^{\frac{1}{2}\left(\Lambda^{(-1)}_{11}-\Lambda^{(-1)}_{22}\right)}&\bigg[ T^{-1}\Lambda^{(-1)}_{12}\sqrt{2\gamma_\nu \mathcal{J}_\nu}\cos\left(\Psi_\nu \right) \\ &+\left(\frac{1}{\epsilon}P_{12}(k,T; \epsilon)-\epsilon\frac{\Lambda^{(-1)}_{12}\Lambda^{(-1)}_{22}}{2T^2} \right)\sqrt{\frac{2\mathcal{J}_\nu}{\gamma_\nu}}\sin\left(\Psi_\nu \right) \bigg]~,\\[5mm]
\end{split} 
\end{equation}

\begin{equation}
    \begin{split}
    \epsilon\,\xi_\nu(k,T; \epsilon) = -\epsilon\, (-T)^{\frac{1}{2}\left(\Lambda^{(-1)}_{22}-\Lambda^{(-1)}_{11}\right)}&\bigg[ T^{-1}\Lambda^{(-1)}_{21}\sqrt{2\gamma_\mu \mathcal{J}_\mu}\cos\left(\Psi_\mu \right)\\ &+\left(\frac{1}{\epsilon}P_{21}(k,T; \epsilon)-\epsilon\frac{ \Lambda^{(-1)}_{11}\Lambda^{(-1)}_{21}}{2T^2}\right)\sqrt{\frac{2\mathcal{J}_\mu}{\gamma_\mu}}\sin\left(\Psi_\mu \right)\bigg]~.
\end{split}  
\end{equation}

Explicitly, the zero-th order components have the following expressions
\begin{subequations}\label{Explicit form of F and G oscillating}
\begin{align}
   \begin{split}
        \hat{F}^{(0)}_\mu=-\frac{\gamma^{(0) \prime}_\mu}{\gamma^{(0)}_\mu}\mathcal{J}^{(0)}_\mu \cos (2\Psi^{(0)}_\mu) + 2(-T)^{\frac{1}{2}\left(\Lambda^{(-1)}_{11}-\Lambda^{(-1)}_{22}-2\right)}\Lambda^{(-1)}_{12}\sqrt{\frac{ \gamma^{(0)}_\nu \mathcal{J}^{(0)}_\mu \mathcal{J}^{(0)}_\nu}{\gamma^{(0)}_\mu}}\cos(\Psi^{(0)}_\nu)\cos(\Psi^{(0)}_\mu) \\ - 2 (-T)^{\frac{1}{2}\left(\Lambda^{(-1)}_{11}-\Lambda^{(-1)}_{22}\right)}P^{(1)}_{12}(k,T; \epsilon)\sqrt{\frac{ \mathcal{J}^{(0)}_\mu \mathcal{J}^{(0)}_\nu}{\gamma^{(0)}_\mu \gamma^{(0)}_\nu }} \cos(\Psi^{(0)}_\mu) \sin(\Psi^{(0)}_\nu)~,
   \end{split}\\
   \begin{split}
        \hat{F}^{(0)}_\nu=-\frac{\gamma^{(0)\prime}_\nu}{\gamma^{(0)}_\nu}\mathcal{J}^{(0)}_\nu \cos (2\Psi^{(0)}_\nu) + 2(-T)^{\frac{1}{2}\left(\Lambda^{(-1)}_{22}-\Lambda^{(-1)}_{11}-2\right)}\Lambda^{(-1)}_{21}\sqrt{\frac{ \gamma^{(0)}_\mu \mathcal{J}^{(0)}_\nu \mathcal{J}^{(0)}_\mu}{\gamma^{(0)}_\nu}}\cos(\Psi^{(0)}_\mu)\cos(\Psi^{(0)}_\nu) \\ - 2 (-T)^{\frac{1}{2}\left(\Lambda^{(-1)}_{22}-\Lambda^{(-1)}_{11}\right)}P^{(1)}_{21}(k,T; \epsilon)\sqrt{\frac{ \mathcal{J}^{(0)}_\nu \mathcal{J}^{(0)}_\mu}{\gamma^{(0)}_\nu \gamma^{(0)}_\mu }} \cos(\Psi^{(0)}_\nu) \sin(\Psi^{(0)}_\mu)~,
   \end{split}\\
    \begin{split}
        \hat{G}^{(0)}_\mu =  \frac{\gamma^{(0)\prime}_\mu}{2\gamma^{(0)}_\mu}  \sin(2\Psi^{(0)}_\mu) -(-T)^{\frac{1}{2}\left(\Lambda^{(-1)}_{11}-\Lambda^{(-1)}_{22}-2\right)}\Lambda^{(-1)}_{12}\sqrt{\frac{\gamma^{(0)}_\nu \mathcal{J}^{(0)}_\nu }{\gamma^{(0)}_\mu \mathcal{J}^{(0)}_\mu}} \sin(\Psi^{(0)}_\mu)\cos(\Psi^{(0)}_\nu) \\ + (-T)^{\frac{1}{2}\left(\Lambda^{(-1)}_{11}-\Lambda^{(-1)}_{22}\right)}P^{(1)}_{12}(k,T;\epsilon) \sqrt{\frac{\mathcal{J}^{(0)}_\nu }{\mathcal{J}^{(0)}_\mu \gamma^{(0)}_\mu \gamma^{(0)}_\nu}} \sin(\Psi^{(0)}_\nu) \sin(\Psi^{(0)}_\mu)~,
    \end{split}\\
    \begin{split}
        \hat{G}^{(0)}_\nu = \frac{\gamma^{(0)\prime}_\nu}{2\gamma^{(0)}_\nu}  \sin(2\Psi^{(0)}_\nu) -(-T)^{\frac{1}{2}\left(\Lambda^{(-1)}_{22}-\Lambda^{(-1)}_{11}-2\right)}\Lambda^{(-1)}_{21}\sqrt{\frac{\gamma^{(0)}_\mu \mathcal{J}^{(0)}_\mu }{\gamma^{(0)}_\nu \mathcal{J}^{(0)}_\nu}} \sin(\Psi^{(0)}_\nu)\cos(\Psi^{(0)}_\mu) \\ + (-T)^{\frac{1}{2}\left(\Lambda^{(-1)}_{22}-\Lambda^{(-1)}_{11}\right)}P^{(1)}_{21}(k,T;\epsilon) \sqrt{\frac{\mathcal{J}^{(0)}_\mu }{\mathcal{J}^{(0)}_\nu \gamma^{(0)}_\nu \gamma^{(0)}_\mu}} \sin(\Psi^{(0)}_\mu) \sin(\Psi^{(0)}_\nu)~.
    \end{split}
\end{align}
\end{subequations}
As discussed in the main text, higher-order expansion coefficients of the oscillating components are not needed for our purposes, as they do not enter the first-order perturbative solution.

The following line integrals arise in the first-order solution \eqref{Eq:fullsolutions_JPsi}
{\allowdisplaybreaks
\begin{subequations}\label{Eq:mixing_terms}
\begin{align}
   \begin{split}
        &\int\de s\;\hat{F}^{(0)}_{\mu}(\mathcal{J}_i^{(0)},\Omega_i s+f_{\Psi_i}^{(0)},T)\Big\lvert_{\Omega_i s=\tau_i}=-\frac{\gamma^{(0) \prime}_\mu}{\gamma^{(0)}_\mu}\mathcal{J}^{(0)}_\mu  \frac{\sin(2\Psi^{(0)}_{\mu})}{2\Omega_\mu}  \\ 
        &+ (-T)^{\frac{1}{2}\left(\Lambda^{(-1)}_{11}-\Lambda^{(-1)}_{22}-2\right)}\Lambda^{(-1)}_{12}\sqrt{\frac{\gamma^{(0)}_\nu \mathcal{J}^{(0)}_\mu \mathcal{J}^{(0)}_\nu}{\gamma^{(0)}_\mu}}\left(\frac{\sin(\Psi^{(0)}_\mu - \Psi^{(0)}_\nu)}{(\Omega_\mu -\Omega_\nu)} +\frac{\sin(\Psi^{(0)}_\mu + \Psi^{(0)}_\nu)}{(\Omega_\mu +\Omega_\nu)} \right) \\ 
        &- (-T)^{\frac{1}{2}\left(\Lambda^{(-1)}_{11}-\Lambda^{(-1)}_{22}\right)}P^{(1)}_{12}(k,T; \epsilon) \sqrt{\frac{\mathcal{J}^{(0)}_\mu \mathcal{J}^{(0)}_\nu}{\gamma^{(0)}_\mu \gamma^{(0)}_\nu}} \left(\frac{\cos(\Psi^{(0)}_\mu - \Psi^{(0)}_\nu)}{(\Omega_\mu -\Omega_\nu)} -\frac{\cos(\Psi^{(0)}_\mu + \Psi^{(0)}_\nu)}{(\Omega_\mu +\Omega_\nu)} \right)~, 
   \end{split}\\
   \begin{split}
        &\int\de s\;\hat{F}^{(0)}_{\nu}(\mathcal{J}_i^{(0)},\Omega_i s+f_{\Psi_i}^{(0)},T)\Big\lvert_{\Omega_i s=\tau_i}=-\frac{\gamma^{(0)\prime}_\nu}{\gamma^{(0)}_\nu}\mathcal{J}^{(0)}_\nu \frac{\sin (2\Psi^{(0)}_{\nu})}{2\Omega_\nu}  \\ 
        &+ (-T)^{\frac{1}{2}\left(\Lambda^{(-1)}_{22}-\Lambda^{(-1)}_{11}-2\right)}\Lambda^{(-1)}_{21}\sqrt{\frac{ \gamma^{(0)}_\mu \mathcal{J}^{(0)}_\nu \mathcal{J}^{(0)}_\mu}{\gamma^{(0)}_\nu}}\left(\frac{\sin(\Psi^{(0)}_\mu - \Psi^{(0)}_\nu)}{(\Omega_\mu -\Omega_\nu)} +\frac{\sin(\Psi^{(0)}_\mu + \Psi^{(0)}_\nu)}{(\Omega_\mu +\Omega_\nu)} \right) \\ 
        &- (-T)^{\frac{1}{2}\left(\Lambda^{(-1)}_{22}-\Lambda^{(-1)}_{11}\right)}P^{(1)}_{21}(k,T; \epsilon) \sqrt{\frac{ \mathcal{J}^{(0)}_\nu \mathcal{J}^{(0)}_\mu}{\gamma^{(0)}_\nu \gamma^{(0)}_\mu 
        }} \left(\frac{\cos(\Psi^{(0)}_\nu - \Psi^{(0)}_\mu)}{(\Omega_\nu -\Omega_\mu)} -\frac{\cos(\Psi^{(0)}_\nu + \Psi^{(0)}_\mu)}{(\Omega_\mu +\Omega_\nu)} \right)~.
   \end{split}\\
    \begin{split}
        &\int\de s\; \hat{G}^{(0)}_\mu(\mathcal{J}_i^{(0)},\Omega_i s+f_{\Psi_i}^{(0)},T)\Big\lvert_{ \Omega_i s= \tau_i} = -\frac{\gamma^{(0) \prime}_\mu}{\gamma^{(0)}_\mu}\frac{\cos(2\Psi^{(0)}_\mu)}{4\Omega_\mu}  \\ 
        &+ (-T)^{\frac{1}{2}\left(\Lambda^{(-1)}_{11}-\Lambda^{(-1)}_{22}-2\right)}\Lambda^{(-1)}_{12}\sqrt{\frac{\gamma^{(0)}_\nu \mathcal{J}^{(0)}_\nu }{\gamma^{(0)}_\mu \mathcal{J}^{(0)}_\mu}} \left( \frac{\cos(\Psi^{(0)}_\mu + \Psi^{(0)}_\nu)}{2(\Omega_\mu +\Omega_\nu)} + \frac{\cos(\Psi^{(0)}_\mu - \Psi^{(0)}_\nu)}{2(\Omega_\mu -\Omega_\nu)}\right) \\ 
        &+ (-T)^{\frac{1}{2}\left(\Lambda^{(-1)}_{11}-\Lambda^{(-1)}_{22}\right)}P^{(1)}_{12}(k,T;\epsilon)\sqrt{\frac{\mathcal{J}^{(0)}_\nu }{\mathcal{J}^{(0)}_\mu \gamma^{(0)}_\mu\gamma^{(0)}_\nu}}\left(\frac{\sin(\Psi^{(0)}_\mu - \Psi^{(0)}_\nu)}{2(\Omega_\mu -\Omega_\nu)} -\frac{\sin(\Psi^{(0)}_\mu + \Psi^{(0)}_\nu)}{2(\Omega_\mu +\Omega_\nu)} \right)~, 
    \end{split}\\
    \begin{split}
       & \int\de s\; \hat{G}^{(0)}_\nu(\mathcal{J}_i^{(0)},\Omega_i s+f_{\Psi_i}^{(0)},T)\Big\lvert_{ \Omega_i s= \tau_i} = -\frac{\gamma^{(0)\prime}_\nu}{\gamma^{(0)}_\nu} \frac{\cos(2\Psi^{(0)}_\nu)}{4\Omega_\nu} \\ 
       &+ (-T)^{\frac{1}{2}\left(\Lambda^{(-1)}_{22}-\Lambda^{(-1)}_{11}-2\right)}\Lambda^{(-1)}_{21}\sqrt{\frac{\gamma^{(0)}_\mu \mathcal{J}^{(0)}_\mu }{\gamma^{(0)}_\nu \mathcal{J}^{(0)}_\nu}} \left( \frac{\cos(\Psi^{(0)}_\nu + \Psi^{(0)}_\mu)}{2(\Omega_\nu +\Omega_\mu)} + \frac{\cos(\Psi^{(0)}_\nu - \Psi^{(0)}_\mu)}{2(\Omega_\nu -\Omega_\mu)}\right) \\ 
       &+ (-T)^{\frac{1}{2}\left(\Lambda^{(-1)}_{22}-\Lambda^{(-1)}_{11}\right)}P^{(1)}_{21}(k,T;\epsilon)\sqrt{\frac{\mathcal{J}^{(0)}_\mu }{\mathcal{J}^{(0)}_\nu \gamma^{(0)}_\nu\gamma^{(0)}_\mu}}\left(\frac{\sin(\Psi^{(0)}_\mu - \Psi^{(0)}_\nu)}{2(\Omega_\mu -\Omega_\nu)} -\frac{\sin(\Psi^{(0)}_\mu + \Psi^{(0)}_\nu)}{2(\Omega_\mu +\Omega_\nu)} \right)~, 
    \end{split}
\end{align}
\end{subequations}
}
For the explicit evaluation of the above expressions, it is useful to recall that the zeroth-order solutions are 
\begin{subequations}
    \begin{align}
        &{\mathcal{J}^{(0)}_N}={\zeta^{(0)}_N}~,\\
         &{\Psi^{(0)}_N}(\tau_N,T)=\tau_N+f^{(0)}_{\Psi_N}~,
    \end{align}
\end{subequations}
with $\tau_\mu=\int_{T_0}^{T}\de u\; \Omega_\mu (u)$\,, $\tau_\nu=\int_{T_0}^{T}\de u \; \Omega_\nu (u)$\,, and ${f^{(0)}_{\Psi_N}}(T)=\int_{T_0}^{T}\de u\;\bar{G}_N^{(0)}(u)$~.

\section{Explicit solutions for the toy models in Section~4}
\subsection{Mass Mixing}
The multiple-scale solutions for the model \eqref{monomial time dependent model} read
{\allowdisplaybreaks
\begin{subequations}\label{Monomial Action and Angle}

    \begin{align}
        \mathcal{J}_\mu &\simeq \zeta^{(0)}_\mu + \epsilon \left(\zeta^{(1)}_\mu - \frac{2^{1/4}m^2}{k^2}\sqrt{\zeta^{(0)}_\mu \zeta^{(0)}_\nu }\left( (2+\sqrt{2})\cos(\Psi^{(0)}_\mu-\Psi^{(0)}_\nu)-(2-\sqrt{2})\cos(\Psi^{(0)}_\mu+\Psi^{(0)}_\nu)\right)\right) ~,  \\
        \mathcal{J}_\nu &\simeq \zeta^{(0)}_\nu + \epsilon \left(\zeta^{(1)}_\nu +\frac{2^{5/4}m^2}{k^2}\sqrt{\zeta^{(0)}_\nu \zeta^{(0)}_\mu }\left( (\sqrt{2}+2)\cos(\Psi^{(0)}_\mu-\Psi^{(0)}_\nu)+(2-\sqrt{2})\cos(\Psi^{(0)}_\mu+\Psi^{(0)}_\nu)\right)\right)~, \\
        \Psi^{(0)}_\mu &= c^{(0)}_\mu+ t -t_0 +\frac{m^2}{4k^2}\left(T^2-T^2_0\right)~, \\ \Psi^{(1)}_\mu &= c^{(1)}_{\mu} -\frac{m^4}{24k^4}(T^3-T_0^3) + \frac{2^{-3/4}m^2}{k^2}\sqrt{\frac{\zeta^{(0)}_\nu}{\zeta^{(0)}_\mu}}\left((2+\sqrt{2})\sin(\Psi^{(0)}_\mu-\Psi^{(0)}_\nu)
        -(2-\sqrt{2})\sin(\Psi^{(0)}_\mu+\Psi^{0}_\nu)\right)~,\\
        \Psi^{(0)}_{\nu} &= c^{(0)}_{\nu}+\frac{t-t_0}{\sqrt{2}} + \frac{m^2}{\sqrt{2}k^2}\left(T^2-T^2_0\right)~,\\
        \Psi^{(1)}_\nu &=c^{(1)}_{\nu} -\frac{\sqrt{2}m^4}{3k^4}\left( T^3 -T^3_0\right) +\frac{2^{1/4}m^2}{k^2}\sqrt{\frac{\zeta^{(0)}_\mu}{\zeta^{(0)}_\nu}}\left((2 +\sqrt{2})\sin(\Psi^{(0)}_\mu - \Psi^{(0)}_\nu) +(2-\sqrt{2})\sin(\Psi^{(0)}_\nu +\Psi^{(0)}_\mu)\right)~.
    \end{align}
\end{subequations}
}

The solutions for the more general model \eqref{polynomial model} read 
\begin{subequations}\label{Polynomial Action and Angle}
    \begin{align}
        \mathcal{J}_\mu &\simeq \zeta^{(0)}_\mu +\epsilon  \left( \zeta^{(1)}_\mu -\frac{m^2(T^5+T^6)}{k^2} \sqrt{2\zeta^{(0)}_\mu\zeta^{(0)}_\nu} \left( 2\cos(\Psi^{(0)}_\mu -\Psi^{(0)}_\nu) -\frac{2}{3}\cos (\Psi^{(0)}_\mu + \Psi^{(0)}_\nu) \right) \right)~,  \\
        \mathcal{J}_\nu &\simeq \zeta^{(0)}_\nu +\epsilon  \left( \zeta^{(1)}_\nu -\frac{2m^2(T^2+T^3)}{k^2} \sqrt{2\zeta^{(0)}_\nu\zeta^{(0)}_\mu} \left(- 2\cos(\Psi^{(0)}_\mu -\Psi^{(0)}_\nu) -\frac{2}{3}\cos (\Psi^{(0)}_\nu + \Psi^{(0)}_\mu) \right) \right)~, \\
        \Psi^{(0)}_\mu &= c^{(0)}_{\mu}+t-t_0  + \frac{1}{2k}(T-T_0) +\frac{m^2}{k^2}\left(\frac{1}{8}(T^4-T^4_0) +\frac{1}{10}(T^5-T_0^5)\right)~, \\
        \Psi^{(1)}_\mu &=c^{(1)}_{\mu}- \frac{1}{8k^2}(T-T_0) - \frac{m^2}{16k^3}(T^4-T^4_0)-\frac{m^2}{20k^3}(T^5-T^5_0) -\frac{m^4}{56k^4}(T^7-T^7_0)-\frac{m^4}{32k^4}(T^8-T^8_0)\nonumber \\&-\frac{m^4}{72k^4}(T^9-T^9_0) - \frac{m^2}{k^2}\sqrt{\frac{2 \zeta^{(0)}_\nu}{\zeta^{(0)}_\mu}}(T^5 +T^6) \left(\sin(\Psi^{(0)}_\mu-\Psi^{(0)}_\nu)+\frac{1}{3} \sin(\Psi^{(0)}_\mu+\Psi^{(0)}_\nu) \right) ~,\\
        \Psi^{(0)}_\nu &=c^{(0)}_{\nu}+ \frac{t-t_0}{2}+\frac{1}{2k}(T-T_0)+\frac{m^2}{k^2}\left(\frac{2}{5}(T^5-T^5_0)+\frac{1}{3}(T^6-T^6_0\right) ~, \\
        \Psi^{(1)}_\nu &=c^{(1)}_{\nu} - \frac{1}{4k^2}(T-T_0) - \frac{2m^2}{5k^3}(T^5-T^5_0)-\frac{m^2}{3k^3}(T^6-T^6_0) -\frac{4m^4}{9k^4}(T^9-T^9_0)-\frac{4m^4}{5k^4}(T^{10}-T^{10}_0)\nonumber \\&-\frac{4m^4}{11k^4}(T^{11}-T^{11}_0) - \frac{2m^2}{k^2}\sqrt{\frac{2 \zeta^{(0)}_\mu}{\zeta^{(0)}_\nu}}(T^2 +T^3) \left(\sin(\Psi^{(0)}_\mu-\Psi^{(0)}_\nu)+\frac{1}{3} \sin(\Psi^{(0)}_\nu+\Psi^{(0)}_\mu) \right)~.
    \end{align}
\end{subequations}

\subsection{Chiral Mixing}
Below are the solutions for model \eqref{Chiral Model},
\begin{subequations}
    \begin{align}
        \mathcal{J}_\mu &\simeq \zeta_\mu^{(0)} +\epsilon\left(\zeta^{(1)}_\mu - \frac{ 2^{1/4}}{kT}\sqrt{\zeta^{(0)}_\mu \zeta^{(0)}_\nu}\left((2+\sqrt{2})\cos(\Psi^{(0)}_\mu-\Psi^{(0)}_\nu) -(2-\sqrt{2})\cos(\Psi^{(0)}_\mu+\Psi^{(0)}_\nu) \right)\right)~,\\
        \mathcal{J}_\nu &\simeq \zeta_\nu^{(0)} +\epsilon\left(\zeta^{(1)}_\nu + \frac{ 2^{1/4}}{kT}\sqrt{\zeta^{(0)}_\nu \zeta^{(0)}_\mu}\left((2+\sqrt{2})\cos(\Psi^{(0)}_\mu-\Psi^{(0)}_\nu) +(2-\sqrt{2})\cos(\Psi^{(0)}_\mu+\Psi^{(0)}_\nu) \right)\right)~, \\
        \Psi^{(0)}_\mu &= c^{(0)}_{\mu}+t-t_0+\frac{3}{4k}\ln (T/T_0)~, \\
        \Psi^{(1)}_\mu &=c^{(1)}_{\mu} +\frac{9}{32k^2}\lf(\frac{1}{T}-\frac{1}{T_0}\rg) +\frac{ 2^{-3/4}}{kT}\sqrt{\frac{\zeta^{(0)}_\nu}{\zeta^{(0)}_\mu}}\left( (2+ \sqrt{2})\sin(\Psi^{(0)}_\mu -\Psi^{(0)}_\nu) - (2-\sqrt{2})\sin(\Psi^{(0)}_\mu + \Psi^{(0)}_\nu ) \right)~, \\
        \Psi^{(0)}_\nu &=c^{(0)}_{\nu}  +\frac{t-t_0}{\sqrt{2}} +\frac{1}{2\sqrt{2}k}\ln(T/T_0) ~,\\
        \Psi^{(1)}_{\nu} &=c^{(1)}_{\nu}+  \frac{1}{8\sqrt{2}k^2}\left(\frac{1}{T}-\frac{1}{T_0}\right)+\frac{2^{-3/4}}{kT}\sqrt{\frac{\zeta^{(0)}_\mu}{\zeta^{(0)}_\nu}}\left( (2+\sqrt{2})\sin(\Psi^{(0)}_{\mu}-\Psi^{(0)}_{\nu}) - (2-\sqrt{2})\sin(\Psi^{(0)}_\nu+\Psi^{(0)}_\mu) \right)~.
    \end{align}
    \label{Chiral Action and Angle}
\end{subequations}

\subsection{Velocity Mixing}
The solutions for model \eqref{velocity mixing model} read
\begin{subequations}
    \begin{align}
       \mathcal{J}_\mu &\simeq \zeta^{(0)}_\mu +\epsilon \left( \zeta^{(1)}_\mu -  2\sqrt{2\zeta^{(0)}_\mu \zeta^{(0)}_\nu}(T^5+T^6)(\cos(\Psi^{(0)}_\mu - \Psi^{(0)}_\nu))-\frac{1}{3}\cos(\Psi^{(0)}_\mu +\Psi^{(0)}_\nu) \right)~,\\
        \mathcal{J}_\nu &\simeq \zeta^{(0)}_\nu +\epsilon \left( \zeta^{(1)}_\nu +  \sqrt{\frac{\zeta^{(0)}_\nu \zeta^{(0)}_\mu}{2}}(T^3+T^4)(\cos(\Psi^{(0)}_{\mu} - \Psi^{(0)}_{\nu})+\frac{1}{3}\cos(\Psi^{(0)}_\nu +\Psi^{(0)}_\mu)\right)~,\\
        \Psi^{(0)}_\mu &= c^{(0)}_{\mu} + t-t_0 ~,\\
        \Psi^{(1)}_\mu &= c^{(1)}_{\mu}+(T^5 +T^6)\sqrt{\frac{2\zeta^{(0)}_\nu}{\zeta^{(0)}_\mu}} \left( \sin(\Psi^{(0)}_\mu - \Psi^{(0)}_\nu) -\frac{1}{3}\sin(\Psi^{(0)}_\mu +\Psi^{(0)}_\nu)\right) ~,\\
        \Psi^{(0)}_\nu &= c^{(0)}_\nu + \frac{t-t_0}{2} ~,\\
       \Psi^{(1)}_\nu &= c^{(1)}_{\nu} +(T^3+T^4)\sqrt{\frac{\zeta^{(0)}_\mu}{8\zeta^{(0)}_\nu}}\left(\sin(\Psi^{(0)}_{\mu}-\Psi^{(0)}_{\nu})-\frac{1}{3}\sin(\Psi^{(0)}_\nu+\Psi^{(0)}_\mu)\right)~.
    \end{align}
    \label{Velocity Action and Angle}
\end{subequations}

\subsection{Friction Mixing}
The solutions for the model \eqref{Friction mixing model} are shown below,

\begin{subequations}
    \begin{align}
         \mathcal{J}_\mu &\simeq \zeta^{(0)}_\mu +\epsilon \left( \zeta^{(1)}_\mu +\alpha\, 2^{-9/4} (-T)^{-1-2\Delta}\sqrt{\zeta^{(0)}_\mu\zeta^{(0)}_\nu}\left((2+\sqrt{2})\sin(\Psi^{(0)}_\mu-\Psi^{(0)}_\nu)+(2-\sqrt{2})\sin(\Psi^{(0)}_\mu+\Psi^{(0)}_\nu)\right)\right)~, \\
         \mathcal{J}_\nu &\simeq \zeta^{(0)}_\nu +\epsilon \left( \zeta^{(1)}_\nu -\alpha\, 2^{-7/4} (-T)^{-1+2\Delta} \sqrt{\zeta^{(0)}_\nu\zeta^{(0)}_\mu}\left((2+\sqrt{2})\sin(\Psi^{(0)}_\mu-\Psi^{(0)}_\nu)+(2-\sqrt{2})\sin(\Psi^{(0)}_\nu+\Psi^{(0)}_\mu)\right)\right)~, \\
         \Psi^{(0)}_\mu &= c^{(0)}_{\mu} + t - t_0~,\\
         \begin{split}
             \Psi^{(1)}_\mu &= c^{(1)}_\mu + \frac{1}{2}\left(\frac{1}{T}-\frac{1}{T_0}\right)\Delta(\Delta-1)\\ &+\alpha\, 2^{-13/4} (-T)^{-1-2\Delta}\sqrt{\frac{\zeta^{(0)}_\nu}{\zeta^{(0)}_\mu}} \left( (2+\sqrt{2})\cos(\Psi^{(0)}_\mu-\Psi^{(0)}_\nu) +(2-\sqrt{2})\cos(\Psi^{(0)}_\mu+\Psi^{(0)}_\nu)\right)~,
         \end{split}
          \\
          \Psi^{(0)}_\nu &= c^{(0)}_{\nu} +\frac{t -t_0}{\sqrt{2}}~,\\
        \begin{split}
             \Psi^{(1)}_\nu &= c^{(1)}_{\nu} + \frac{1}{\sqrt{2}}\left(\frac{1}{T}-\frac{1}{T_0}\right)\Delta(\Delta+1) \\ + &\alpha\, 2^{-11/4} (-T)^{-1+2\Delta}\sqrt{\frac{\zeta^{(0)}_\mu}{\zeta^{(0)}_\nu}}\left((2+\sqrt{2})\cos(\Psi^{(0)}_{\mu}-\Psi^{(0)}_\nu)-(2-\sqrt{2})\cos(\Psi^{(0)}_{\mu}+\Psi^{(0)}_{\nu})\right)~.
        \end{split}
         \end{align}
         \label{Friction Action and Angle}
\end{subequations}
\newpage

\section{Residual plots for the models in Section 4}\label{Sec:AppendixResidualPlots}

In this Appendix, we show the plots of the residuals (that is, the absolute difference between the multiple-scale and numerical solutions) for the toy models considered in Section \ref{Sec:pheno_model}.

\subsection{Mass Mixing}
\begin{figure}[!htb]
    \centering
    \includegraphics[width=0.9\linewidth]{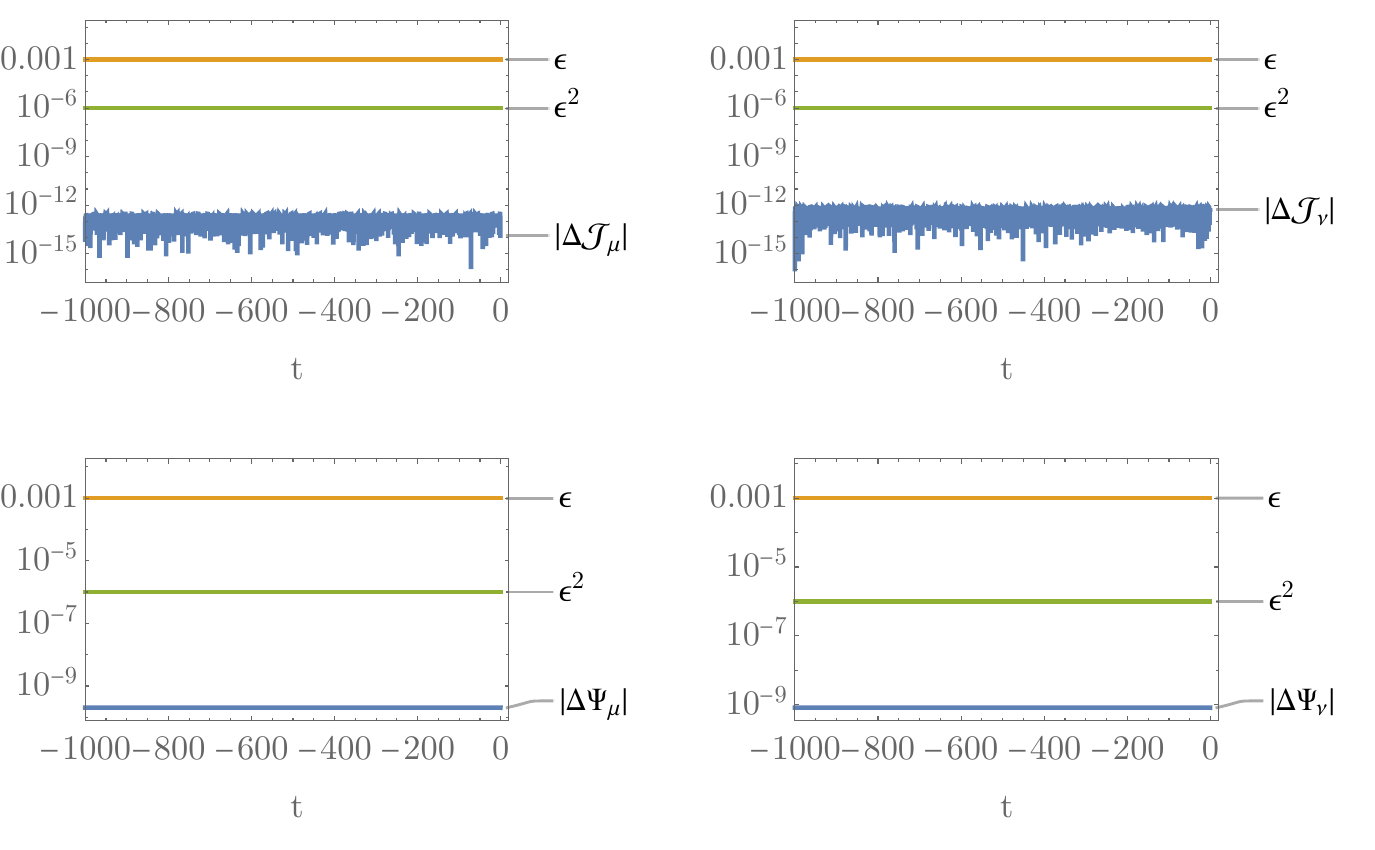}
    \caption{The residuals of the action and angle variables for the model in Eq.~\eqref{monomial time dependent model}. The parameters chosen are the same as in Fig.~\ref{Mass Mixing Monomial time Dependence Late Time}.}
    \label{Monomial Residuals}
\end{figure}

\begin{figure}[!htb]
    \centering
    \includegraphics[width=0.9\linewidth]{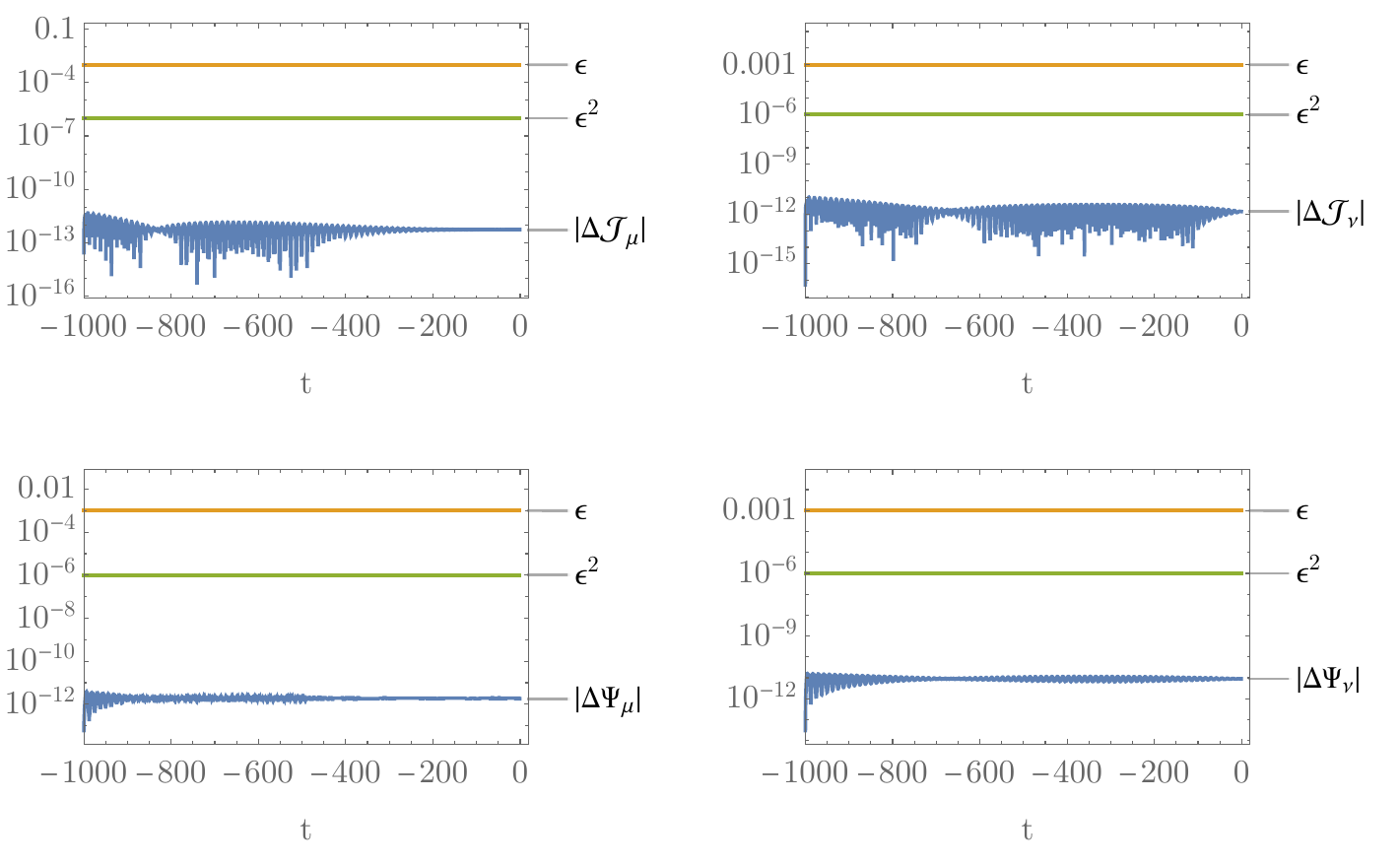}
    \caption{The residuals of the action and angle variables for Eq.~\eqref{polynomial model}. The parameters chosen are the same as stated in Fig.~\ref{Time Dependent Model}.}
    \label{Polynomial residuals}
\end{figure}
\FloatBarrier
\subsection{Chiral Mixing}
\begin{figure}[!htb]
    \centering
    \includegraphics[width=0.9\linewidth]{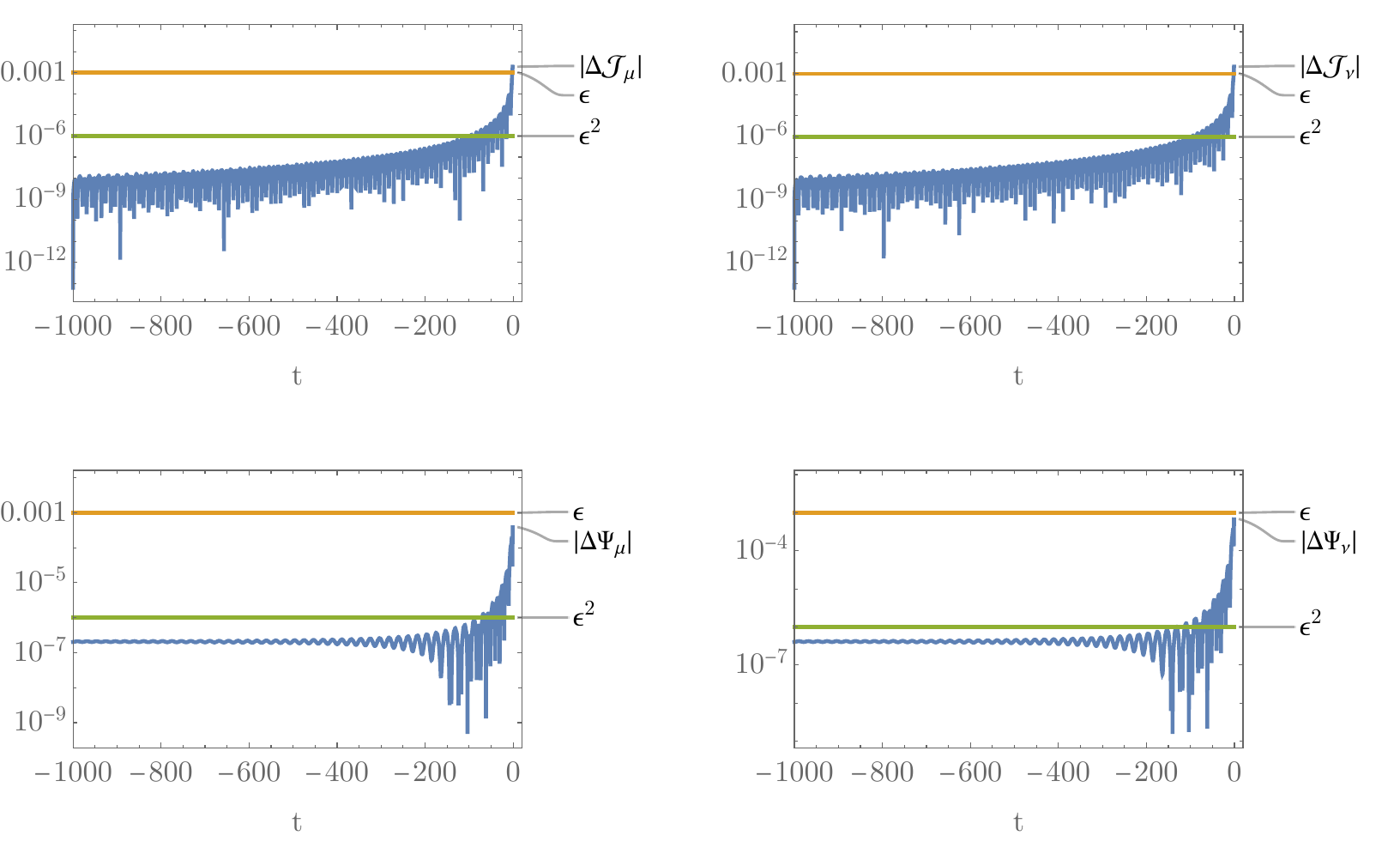}
    \caption{The residuals of the action and angle variables for Eq.~\eqref{Chiral Model}. The parameters chosen are the same stated in Fig.~\ref{Chiral Mixing Toy Model}.}
    \label{Chiral Residuals}
\end{figure}
\FloatBarrier
\subsection{Velocity Mixing}
\begin{figure}[!htb]
    \centering
    \includegraphics[width=0.9\linewidth]{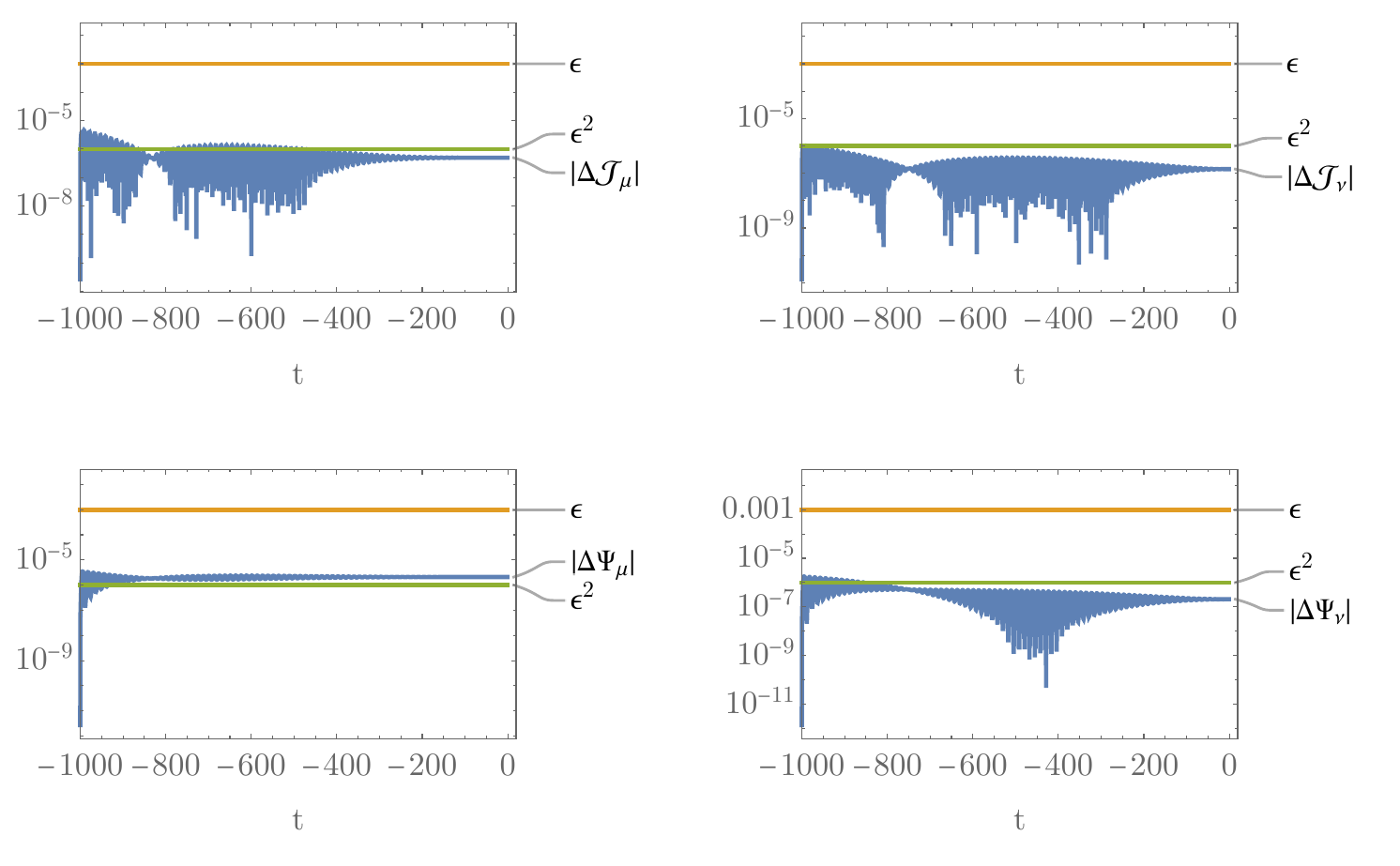}
    \caption{The residuals of the action and angle variables for Eq.~\eqref{velocity mixing model}. The parameters chosen are the same as stated in Fig.~\ref{velocity mixing plots}.}
    \label{Velocity Residuals}
\end{figure}
\FloatBarrier
\subsection{Friction Mixing}
\begin{figure}[!htb]
    \centering
    \includegraphics[width=0.9\linewidth]{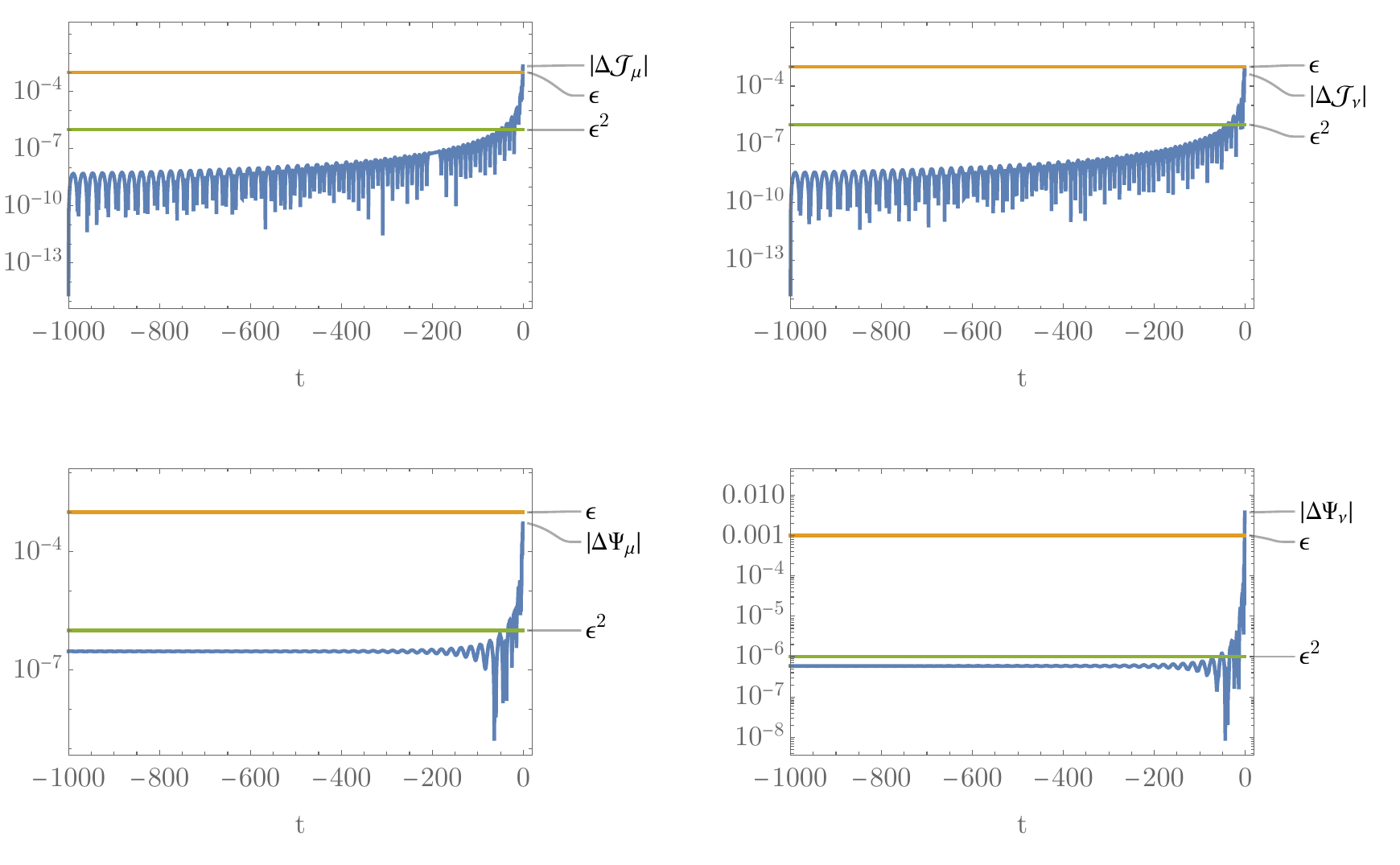}
    \caption{The residuals of the action and angle variables for Eq.~\eqref{Friction mixing model}. The parameters chosen are the same as stated in Fig.~\ref{Friction mixing plots}.}
    \label{Friction Residuals}
\end{figure}
\FloatBarrier

\bibliography{mybib}

\end{document}